\documentclass[a4paper]{article}
\usepackage[margin=2.5cm]{geometry}
\usepackage{caption3} 
\usepackage{hyperref}
\DeclareCaptionOption{parskip}[]{}
\usepackage[small]{caption}
\usepackage{appendix}
\usepackage[pdftex]{graphicx}
\usepackage{natbib}
\usepackage{latexsym}
\usepackage{tabularx}
\usepackage[english]{babel}
\usepackage[latin1]{inputenc}
\usepackage{verbatim}
\usepackage{amsmath}
\usepackage{amssymb}
\usepackage{cleveref}
\usepackage{fancyhdr}
\begin{document}
%opening
\title{Meridional trapping and zonal propagation of inertial waves in a rotating fluid shell}

\author{Anna Rabitti$^1$%
  \thanks{Email address for correspondence: anna.rabitti@nioz.nl. \textbf{Citation:} Rabitti, A., and L. R. M. Maas. 2013. ``Meridional Trapping and Zonal Propagation of Inertial Waves in a Rotating Fluid Shell.'' Journal of Fluid Mechanics 729 (July 24): 445-470. doi:10.1017/jfm.2013.310.} and
Leo R. M. Maas$^1$\\
\textit{\small{NIOZ Royal Netherlands Institute for Sea Research,}}\\
\textit{\small{Department of Physical Oceanography, P.O. Box 59, 1790 AB Den Burg, Texel, The Netherlands}}
}
\date{}

%%%%%%%%%%%%%%%%%%%%%%%%%%%%%%%%%%%%%%%%%%%%%%%%%%%%%%%%%%%%%%%%
\maketitle

\begin{abstract}
Inertial waves propagate in homogeneous rotating fluids, and constitute a challenging and simplified case study for the broader class of inertio-gravity waves, present in all geophysical and astrophysical media, and responsible for energetically costly processes as diapycnal and angular momentum mixing.
However, a complete analytical description and understanding of internal waves in arbitrarily shaped enclosed domains, such as the ocean, or a planet liquid core, is still missing. 

In this work, the inviscid, linear inertial wave field is investigated by means of three dimensional ray tracing in spherical shell domains, having in mind possible oceanographic applications. Rays are here classically interpreted as representative of energy paths. But in contrast with previous studies, they are now launched with a non-zero initial zonal component allowing for a more realistic, localized forcing, and the development of azimuthal inhomogeneities.

We find that meridional planes generally act in the shell geometry as attractors for ray trajectories. In addition, the existence of trajectories that are not subject to meridional trapping is here observed for the first time. Their dynamics was not captured by the previous purely meridional studies and unveils a new class of possible solutions for inertial motion in the spherical shell.

Both observed behaviours shed some new light on possible mechanisms of energy localization, a key process that still deserves further investigation in our ocean, as well as in other stratified, rotating media.
\end{abstract}

\section{Introduction}
\label{sec:introduction}
Internal waves are ubiquitously present and of great importance in astrophysical and geophysical fluids as present in stars or planetary atmospheres and oceans, where they are considered responsible for a substantial part of the dynamics and mixing in the interior of the supporting medium \citep{Wunsch2004,Ogilvie2004}.
They propagate in all kinds of stratified fluids, and are generally referred to as \textit{internal} (\textit{gravity}) waves when the stratification in the fluid is built by means of vertical density changes, and as \textit{inertial} waves when the supporting stratification is in angular momentum (homogeneous, rotating fluid). 
In this paper, we will use the generic term \textit{internal} waves to refer to both situations, or to a combination of the two, stressing that we deal with a propagating perturbation whose maximum amplitude occurs in the interior of the fluid domains, instead of at the boundary, as for surface waves.
Improving our theoretical understanding of internal wave behaviour in stratified fluids naturally confined to enclosed domains bears, as we may expect, important consequences for our general understanding of ocean or atmosphere dynamics. 

Despite their broad interest and fundamental character, the nature of internal waves in confined domains is still largely unknown due to a variety of difficulties that undermine the study of these oscillations. For example, mathematical difficulties arise because of the combination of the hyperbolic character of internal waves (or of the mixed elliptic-hyperbolic nature in the case of a stratified and rotating fluid) and the confinement of motion to an enclosed domain. This problem, in particular in a spherical shell, has been already presented as a paradigmatic mathematically ill-posed Poincar\'{e} problem \citep{Rieutord2000}, and it is limiting our capabilities of an analytical representation and comprehension of the mechanisms involved. 

Analytical solutions for internal wave problems in enclosed domains are mostly known only for exceptionally symmetric geometries (such as the sphere, see \citet{Bryan1889}), in which they lead to regular solutions. For this reason, ray theory \citep{Whitham1974,Broutman2004a} has been developed to study propagation of internal waves in arbitrarily shaped enclosed geometries. 
Ray theory is based on the simple observation that in uniformly stratified, or uniformly rotating fluids, internal waves propagate along beams whose direction is set by the ratio of the perturbing frequency and a frequency representing the environmental conditions (rotation and/or density stratification) (see \S\ref{sec:method} for derivation). 
Wave energy travels along these beams parallel to group velocity \citep{Harlander2006, Harlander2007bb}, which can therefore provide a (partial) view on the energy distribution in the domain and on the regularity of the associated wave field in cases where no analytical solutions are available. 
It is worth noting that each different inclination of the rays represents one frequency, and thus only linear phenomena, characterized by a single frequency, can be described by means of ray tracing studies. 
Because of its geophysical and astrophysical relevance, a two-dimensional meridional cut of the spherical shell geometry (an annulus) has been largely explored by means of this technique \citep{Hughes1964,Bretherton1964,Dintrans1999, Maas2007my}, and the pathological character of its inviscid solutions has been established \citep{Stewartson1969, Stewartson1971, Stewartson1972}.
Singular, discontinuous solutions have indeed to be expected in hyperbolic systems. These interesting features of internal waves in enclosed domains, the so called internal wave attractors \citep{Maas1995}, are easily captured by ray tracing analysis: ray trajectories in arbitrarily shaped containers are generally not closed but, by repeated reflections from the boundaries, converge to a limit cycle (the attractor), in which all the wave energy is concentrated \citep{Maas2005my}. 
Remarkably, in a two-dimensional framework, ray tracing leads to exact, yet geometrically constructed solutions of the inviscid, uniformly stratified fluid equations in arbitrarily shaped fluid domains. Only one exceptional case is known (that of a square-shaped attractor in a trapezoidal domain) where the singular solution is also expressed in terms of a Fourier series \citep{Maas2009a}.
The energy scars left in the domain by attractors are observed in quasi two-dimensional laboratory demonstrations both in density stratified fluids as well as in homogeneous rotating fluids \citep{Maas1997, Maas2001, Manders2003, Hazewinkel2008a}, and agree with two-dimensional numerical simulations \citep{Hazewinkel2010}.

High resolution, weakly viscous numerical experiments performed in astrophysical contexts \citep{Dintrans1999, Rieutord2001, Ogilvie2004, Ogilvie2005, Calkins2010} have brought some new insight on the occurrence of these attractors, in particular in the meridional cut of spherical shell geometries, showing a remarkable agreement with trajectories evaluated using a simple linear, geometrical ray tracing of the characteristics.

Interestingly, in literature, investigation of the spherical shell geometries by means of ray tracing  has always been limited to rays constrained to meridional planes only (any plane containing the rotation axis, gravity and the geometrical centre of the domain). This is motivated by the symmetry of the problem in the azimuthal coordinate \citep{Bryan1889, Friedlander1982a, Friedlander1982b, Dintrans1999}, present both in the domain's geometrical shape, as well as in the forcing mechanisms usually studied (tidal forcing, libration of the inner sphere).
If in an astrophysical framework this approach seems natural and representative of the most common perturbations in the fluid, this appears less obvious when we regard ocean phenomena. 
The presence of meridional boundaries (continents) limits the assumption of axisymmetry of the domain; moreover, monochromatic (e. g. tidal), point-like sources (local storms, local conversion of barotropic into baroclinic tide on strong topographic features) play a role in the dynamics. 
Thus, in principle, and especially if we are interested in the near field response to a forcing, there is no compelling reason why a single perturbation, locally  and anisotropically forced, should propagate in a two-dimensional meridional plane only, and, in this way, lose the possibility to show the occurrence of any zonal inhomogeneities.

For this reason the aim of the present study is to extend the use of the ray tracing technique for internal wave characteristics to fully three-dimensional spherical shell domains, allowing also for zonal propagation.
In this work we restrict ourselves to the study of pure \textit{inertial} waves, while the role of density stratification, and the combination of the two mechanisms, will be briefly discussed at the end of the paper. 
Results for a homogeneous rotating shell are modified locally when radial stratification is present, where curved rays replace the straight characteristics considered here and where turning surfaces may limit the part of the fluid domain accessible to waves \citep{Friedlander1982a, Friedlander1982b, Dintrans1999}; on the other hand, the pure inertial problem summarizes many of the difficulties and the features of the gravito-inertial problem, and findings in this partial case are supposed to have general validity. 

The paper is built as follows: the three-dimensional geometrical ray tracing technique and its application to the case of the shell are described in  \S\ref{sec:method}. 
In \S\ref{sec:results} results of the application of the methodology are presented; the main outcomes consist in (1) the general occurrence of meridional attracting planes, where ray trajectories (energy paths) eventually converge, even when initially launched with a zonal component, (2) the existence of exceptional trajectories, representing waves that are not subject to meridional trapping that reflect endlessly around the domain. Clearly the dynamics of these waves is not captured by the classical, purely meridional approach and unveils a new class of solutions. 
Consequences of attracting planes and edge waves are then qualitatively discussed in \S\ref{sec:discussion}, with special attention to the possible oceanographic implications. 
 
\section{Inertial wave three-dimensional ray tracing}
\label{sec:method}
\subsection{Governing equations}
In a Cartesian ($x,y,z$) reference frame (see figure \ref{fig:reflection}a) the pressure field, $p$, of linear inertial waves in a homogeneous, inviscid, uniformly-rotating, Boussinesq fluid is conveniently described by the Poincar\'{e} equation \citep{Cartan1922}
\begin{equation}
p_{xx}+p_{yy}-\frac{1-\omega^{2}}{\omega^{2}}p_{zz}=0.
\label{eq:PE}
\end{equation}
where the fluid is rotating at rate $\Omega$ oriented parallel to the $z$-direction and subscripts denote partial derivatives. All fields are assumed to be proportional to $e^{i\omega t}$ and the Coriolis parameter $f=2\Omega$ is taken as the characteristic time scale. Frequencies $\omega$ are thus regarded as normalized with respect to this parameter.
Propagating inertial waves exist in the frequency range $0 < \omega < 1$, therefore the Poincar\'{e} equation is hyperbolic throughout the whole domain.

The momentum equations relate the velocity field to spatial gradients of the pressure:
\begin{equation}
\left. \begin{array}{ll}  
\displaystyle u =\frac{1}{\rho_0}\frac{1}{\omega^2-1}(i\omega p_x+p_y)\\[8pt]
\displaystyle v=\frac{1}{\rho_0}\frac{1}{\omega^2-1}(-p_x+i\omega p_y)\\[8pt]
\displaystyle  w= \frac{1}{\rho_0}\frac{i}{\omega}p_z
 \end{array}\right.
 \label{eq:uvw}
\end{equation}
where $\rho_0$ is the density of the fluid. For an inviscid fluid its boundaries are impermeable, requiring vanishing of the normal velocity component at the outer sphere ($x^2 +y^2+ z^2 =1$) and at the inner sphere ($ x^2+y^2 + z^2 = \eta^2$). Here $\eta=r_{in}/r_{out}$, $0 < \eta < 1$, represent the ratio of the radii of the inner and outer shell, constituting the surface and the bottom of our idealized ocean respectively. Spatial variables are rescaled by using $r_{out}$ as length scale.
The pressure field therefore has to obey oblique derivative boundary conditions, which in general prohibits finding exact analytical solutions. 
Some indications of the behaviour of internal waves in such a domain can be obtained, as anticipated, by means of three-dimensional ray tracing in the fluid gap.

\subsection{Geometrical mapping} 
\label{sec:geom}
In this section we will first motivate the use of three-dimensional ray tracing, together with the assumptions underlying the application of this methodology to the problem. 
The horizontal scattering of a single inertial wave ray is then presented at the end of the section as a map relating a reflected to an incident ray.
The interested readers can refer to appendix \ref{app:shell_algo} for the map construction and the computational details.

After horizontally reorienting the $x$-axis along the propagating direction of a single ray, the system can then be described in two-dimensions only, one vertical dimension and one horizontal dimension. 
Substituting a wave like solution $p=e^{i(kx+mz)}$ in the appropriately rotated version of equation (\ref{eq:PE}), the dispersion relation follows:
\begin{equation}
 \omega^2 = \frac{m^2}{k^2+m^2}
\label{eq:dispersion}
\end{equation}
that can be rewritten as
\begin{equation}
 \omega^2 = \sin^2\theta.
\label{eq:dispersion2}
\end{equation}
where the wave vector is now expressed in polar coordinates as $\vec{k}=\kappa(\cos\theta,\sin\theta)$, with $\theta$ the angle between the wave vector and the horizontal.
From equation (\ref{eq:dispersion}) and (\ref{eq:dispersion2}) it follows that for internal waves, group velocity ($\vec{c}_{g}=\nabla_{\vec{k}}\omega$) is perpendicular to phase velocity ($\vec{c}=\frac{\omega}{\kappa^2}\vec{k}$), and forms an angle $\theta$ with the vertical.
Moreover the factor preceding the second vertical derivative term in equation (\ref{eq:PE}) is directly related to the ray's (group velocity's) inclination since:
\begin{equation}
 \frac{1-\omega^2}{\omega^2} =\cot^2 \theta.
\label{eq:dispersion3}
\end{equation}
This explains why, differently from surface waves, internal wave group velocity in stably stratified fluids is directed along beams, the internal wave rays, whose inclination with respect to the restoring force is uniquely set by the frequency of the perturbation, and the environmental condition (density stratification and/or rotation) \citep{Gortler1943, Greenspan1968}, and this direction is conserved upon reflection at the domain's boundaries.
Here, in analogy with the two-dimensional counterpart, we assume that in a three-dimensional domain, a uniquely connected set of characteristics (rays) exists. Therefore the perturbation will travel now along characteristic \textit{cones} (given by the $2\pi$ rotation of the classical St. Andrew's cross), whose aperture $2\theta$, centred at the rotation axis, is uniquely set by the perturbation frequency and the stratification properties.

Tracing the geometrical trajectories of these rays in order to infer properties of the wave field provides, of course, just a limited perspective on the phenomenon.
In parallel to the approach in geometrical optics for linear problems, in order to successfully apply ray tracing in the three-dimensional case, the propagating perturbation is idealized as a short, plane wave, reflecting on the curved boundary as if from the local tangent plane (see also \citet{Baines1971}).
It is clear that plane waves imply an infinite lateral extension, and its neglect of the (locally) curved character of the spherical domain's boundary constitutes a severe approximation. However, it is only under this strong (as well as popular, e.g. \citet{Whitham1974}) assumption that the geometrical effects can be isolated and studied from the whole complicated excited wave field.
It is remarkable though that the presence of an attractor strengthens the validity of the short wave hypothesis, consistently reducing the width of a wave beam (wave length) while it gets focussed onto the limit cycle.

Now consider we want to trace the behaviour of a perturbation at definite frequency $\omega$. 
In literature, the behaviour of a ray in the spherical shell domain has been traditionally inferred by employing a rotational symmetry of its two-dimensional trajectory, which is therefore traced on a meridional cut of the shell domain solely.
By contrast, in this work, a single ray will be followed in its fully three-dimensional trajectory, while it bounces around the domain.
The perturbation is ``launched'' at one location on the boundary, at position $\vec{x}_0=(x_{0}, 0, \sqrt{\eta^2-x_0^2})$, on the inner sphere (bottom), or at $(x_{0}, 0, \sqrt{1-x_0^2})$, on the outer sphere (surface).
It is clear that we can choose $y_0\equiv0$ because of axial symmetry of basin and equations.
We will then be able to trace one single ray at a time, and it will be uniquely defined by three parameters: $\omega$, $\vec{x}_0$ and its initial horizontal direction $\phi_{0}$, measured anticlockwise with respect to the $x$-axis, which distinguishes it from other rays belonging to the same excited internal wave cone.

Horizontal scattering of the ray from a reflecting boundary is assumed here to be as simple as possible, the adopted scheme being the same as in \citet{Phillips1963} and in \citet{Hughes1964}, already successfully and repetitively used by \citet{Manders2004}. 
Since the wave frequency does not change, also the beam's angle with respect to the vertical will not change upon reflections at the boundaries: this is equivalent to requiring that the incident and reflected waves obey the same dispersion relation, while the boundary condition of vanishing normal flow at the reflection point is also always satisfied.
The mechanism of horizontal scattering of the ray is sketched in figure \ref{fig:reflection}, and explained in detail in appendix \ref{app:shell_algo}.
\begin{figure}
\begin{minipage}[c]{0.2\linewidth}
    \includegraphics{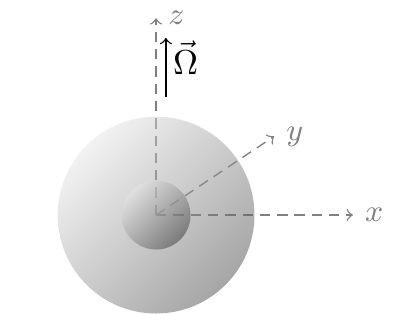}
   \end{minipage}
\begin{minipage}[c]{0.4\linewidth}
    \includegraphics{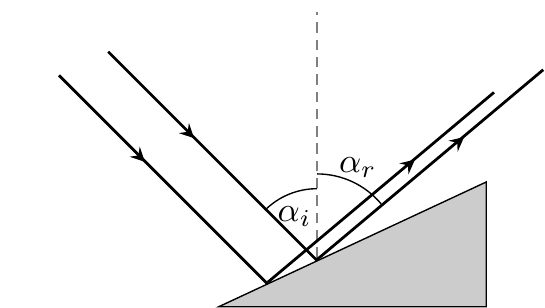}
      \end{minipage}
\begin{minipage}[c]{0.4\linewidth}
    \includegraphics{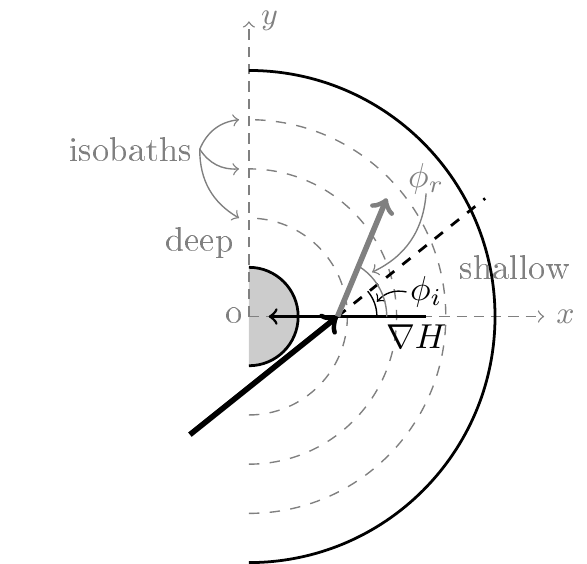}
    \end{minipage}\\

\begin{minipage}[c]{0.2\linewidth}
   ~
   \end{minipage}
\begin{minipage}[c]{0.4\linewidth}
    \centering{side view}
      \end{minipage}
\begin{minipage}[c]{0.4\linewidth}
  \centering{top view}
    \end{minipage}
\caption{(a) Rotating fluid shell inserted in the Cartesian coordinate framework, (b) Side and (c) top view of a short internal wave packet that reflects subcritically from a sloping bottom. In (c) the rotation axis (coinciding with the $z$-axis) is pointing towards the reader. Modified after \citet{Maas2005my}. While incident and reflected rays lie on the same cone (whose angle $\theta$ with the vertical is fixed by the perturbing frequency), their angle $\alpha_i$ and $\alpha_r$ differ in a projection on the vertical plane perpendicular to the slope at the point of reflection. Reflecting waves refract instanteously changing the horizontal propagation direction from $\phi_i$ to $\phi_r$. \label{fig:reflection}}
\end{figure}
While the wave vector component in the along-slope, tangential direction is unchanged, the wave vector component in the cross-slope direction changes due to a focusing or defocusing reflection.
The new horizontal direction ($\phi_r$) of the reflected ray is in fact completely determined by the local bottom slope $s = |\nabla H|$ and horizontal direction of the incoming wave, $\phi_i$ (which equals $\phi_0$ plus the angle that the bottom gradient vector makes with the $x$-direction). 
Conservation and geometrical laws \citep{Phillips1963, Eriksen1985, Gilbert1989, Manders2004, Maas2005my} yield the following relation between $\phi_i$ and $\phi_r$, after the vertical has been correctly stretched to maintain $\tan\theta = 1 $ (see appendix \ref{app:shell_algo} for derivation):
\begin{equation}
 \sin \phi_r = \frac{(s^{2}-1)\sin\phi_i}{2s\cos\phi_{i}+s^{2}+1}.
\label{eq:phir}
\end{equation}

When the reflection takes place where the local bottom slope is smaller than the inclination of the ray ($s<1$), the reflection is called subcritical, and it leads to a change in sign of the vertical component of the ray's group velocity. In the opposite scenario ($s>1$), the reflection is called supercritical, and no change in sign is involved. 
Critical lines (latitudes) connect critical points at which the bottom slope equals the ray slope ($s=1$), and they lead to an exceptional reflection, when non-linear effects are likely to come into play \citep{Dauxois1999, Thorpe1997}. 

The inertial wave ray path for a given frequency $\omega$ and initial launching position and direction ($\vec{x}_0,\phi_0$) can thus be followed as it bounces through the spherical gap applying the known reflection laws and computing subsequent reflection points, each characterized by $\omega,\vec{x}_n,\phi_n $. 
The behaviour of the ray path will provide us information about the features of the unknown solution of the corresponding Poincar\'{e} equation. 

In two-dimensional frameworks, it appears that the existence or absence of eigenmodes in a system is related to the behaviour of these rays and their reflections from the boundaries of the domain: when each characteristic is closed, eigenmodes exist (although infinitely degenerate, \citep{Munnich1996}). In more generic cases, when limit trajectories (attractors) arise, these are the signature of singular, discontinuous field solutions, as shown for the spherical shell \citep{Harlander2007bb}.

A universal, meaningful relation between rays, energy paths and characteristics in three dimensional domains is not completely established yet.
The existence of a similar relation between wave rays and wave paths in three dimensional geometries is supported by recent numerical work \citep{Drijfhout2007} and in laboratory experiments, performed in a non-centrally forced paraboloidal basin \citep{Hazewinkel2010}, but definitely deserves further investigation.

A need for three-dimensional ray tracing also emerges from the work by \cite{Rieutord2010}, where the discrepancies between three-dimensional numerical simulations, analytical solutions, and the two-dimensional, meridional ray orbits, clearly show the limit of the latter approach.
 
\section{Results}
\label{sec:results}
\subsection{Summary of established results for meridional ray motion ($\phi_0=0,\pi$)}
\label{sec:meridional}
\begin{figure}
\centerline{
a)
 \includegraphics[width=.3\linewidth]{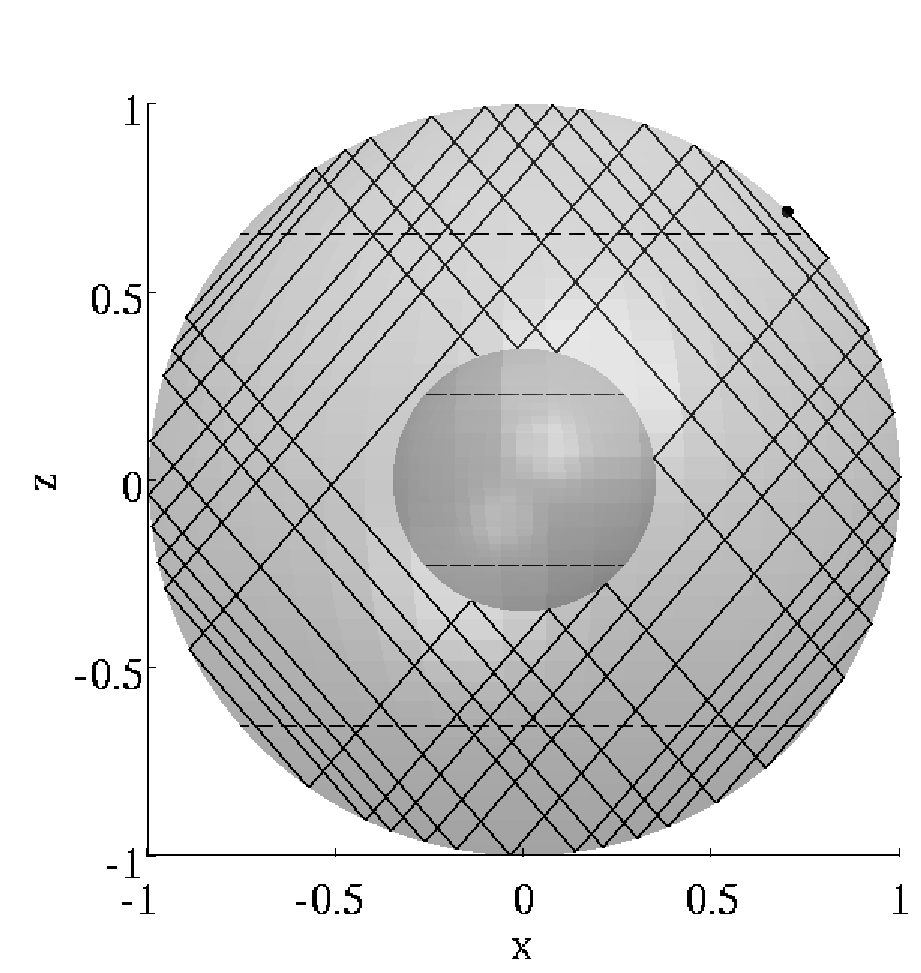}
b)
 \includegraphics[width=.3\linewidth]{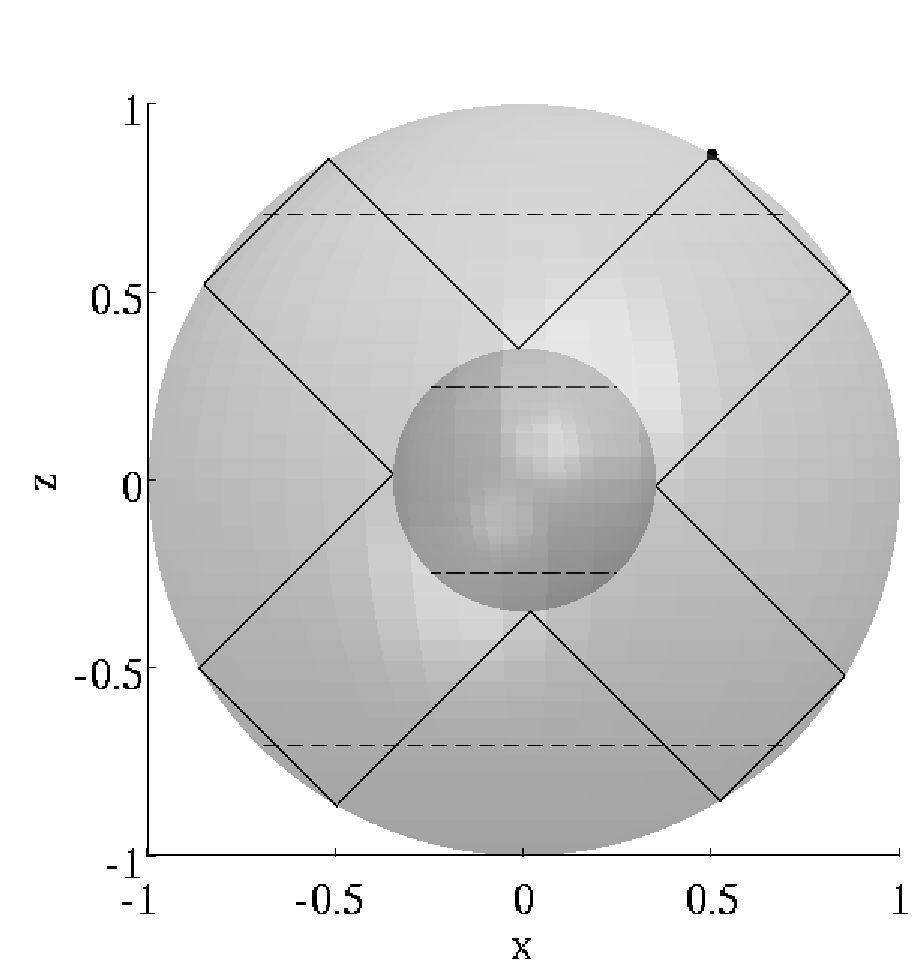}
c)
 \includegraphics[width=.3\linewidth]{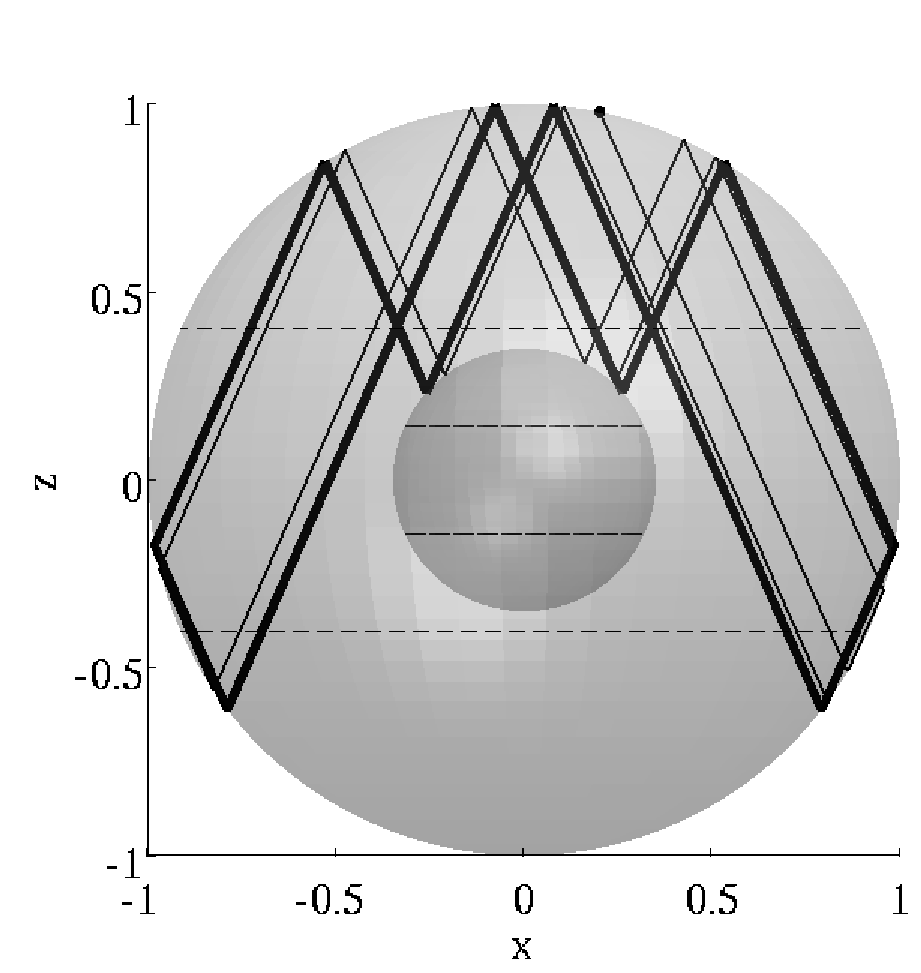}}
\caption{Examples of purely meridional trajectories: (a) ergodic-like orbit, obtained with parameter $\omega = \sqrt{\frac{3}{7}}$, $x_0 = 0.7$ , (b) periodic orbit, $\omega = \sqrt{\frac{1}{2}}$, $x_0 = 0.5$ and (c) attractive orbit, $\omega = 0.4051$, $x_0 = 0.2$. For all cases $\eta = 0.35$. Horizontal lines correspond to critical latitudes. Black dot corresponds to the launching position $x_0$. In (c) the thick line indicates the wave attractor. \label{fig:2d_behaviour}}
\end{figure}
\begin{figure}
\centerline{
a)
\includegraphics[width=.5\linewidth]{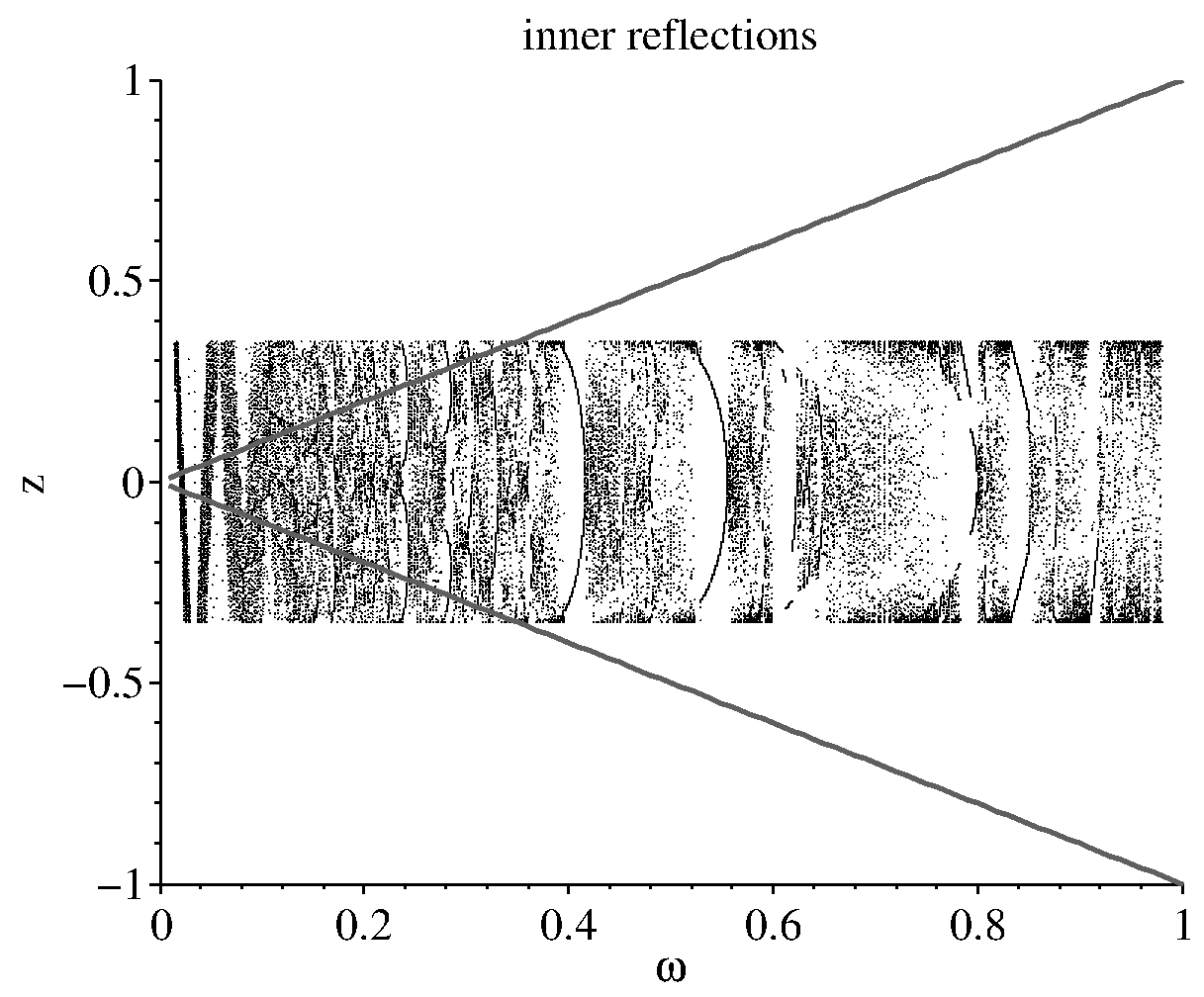}
b)
\includegraphics[width=.5\linewidth]{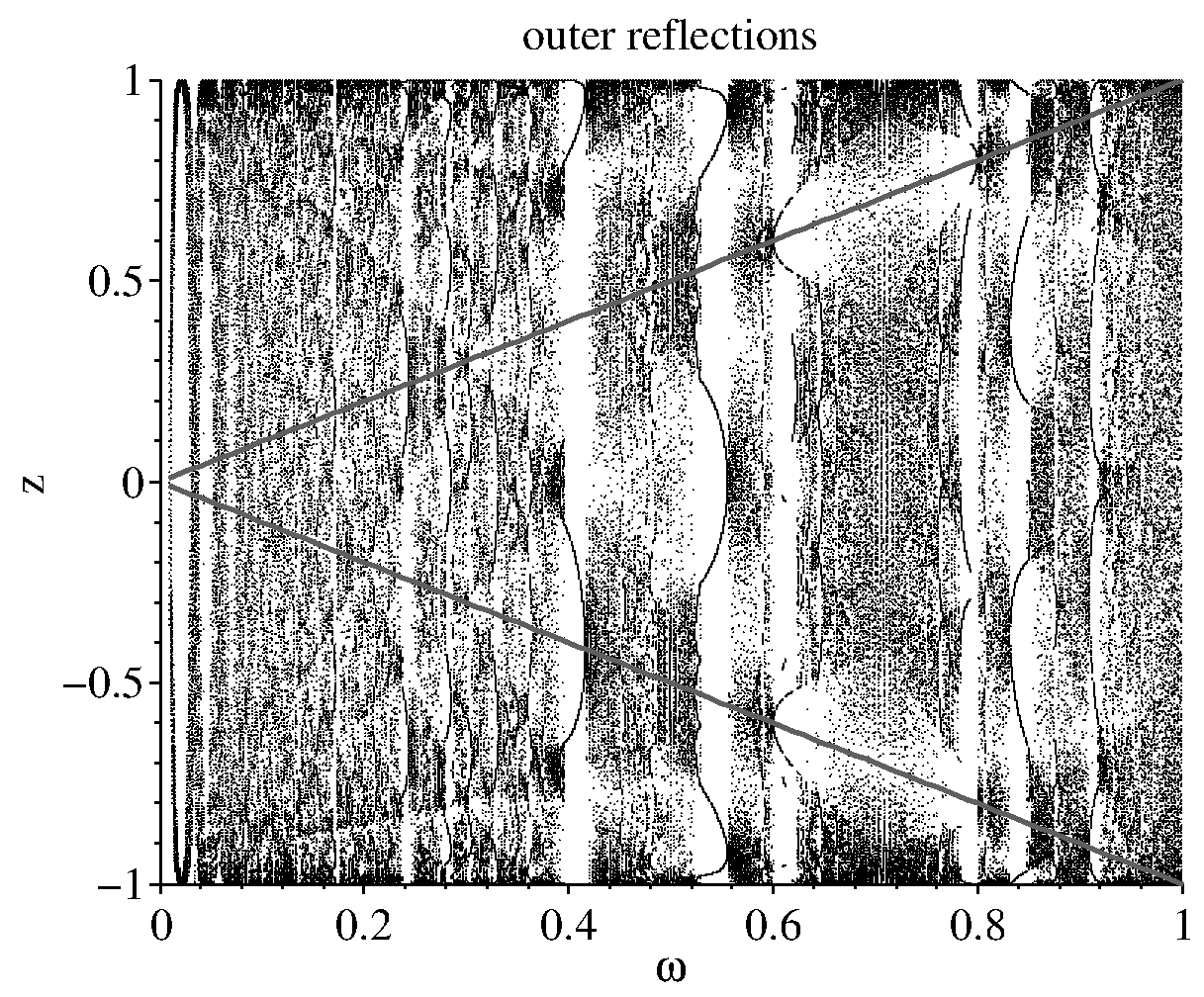}}
\caption{Poincar\'{e} plots for trajectories confined in a meridional section of a spherical shell ($\eta = 0.35$, $x_0=0.2$). On the horizontal axis, $\omega$ is running from $0.01$ to $1$ in steps of $0.0001$. On the  vertical axis, the $z$ coordinate of reflection points on the inner sphere (a) and on the outer sphere (b) of the last 20 reflections (out of 200). Grey lines represent the locations of the critical latitudes.\label{fig:2d_poincare}}
\end{figure}

It is known that in two-dimensional domains, such as a meridional section of a spherical shell (an annulus), ray trajectories show three possible kinds of behaviour \citep{John1941, Maas1995, Dintrans1999}: (1) a single (or denumerable set of) orbit(s) can be plane filling (ergodic, figure \ref{fig:2d_behaviour}a), and represents an annihilating solution, (2) each orbit can be periodic (close onto itself after a number of reflections), representing regular solutions in the domain (figure \ref{fig:2d_behaviour}b), or (3) each orbit can eventually be trapped on one (or a denumerable set of) limit cycle(s), attractor(s), representing singular solution(s) (figure \ref{fig:2d_behaviour}c), which may include point attractors.
In this annulus, periodic two-dimensional orbits are found when the critical latitude ($\lambda_{c}$), where boundary slope equals ray slope, is commensurable with $\pi$. For example, in the case depicted in figure \ref{fig:2d_behaviour}b, $\omega= \sqrt{\frac{1}{2}}$ and $\lambda_{c} = \pi/4$ \citep{Rieutord2001}.

Attractors are found in frequency windows in the inertial range $0<\omega<1$, as shown in \citet{Rieutord2001} and \citet{Maas2007my}. In figure \ref{fig:2d_poincare}, a typical Poincar\'{e} plot for a meridional section of a spherical shell ($\eta = 0.35$) is presented, where frequency is on the horizontal axis, and the vertical axis shows $z$ coordinates of the last $20$ (out of $200$) reflections on the inner (a) and on the outer sphere (b).
As noticed by \citet{Maas1995}, critical latitudes seem to act as repellors for ray trajectories, especially at the outer sphere (figure \ref{fig:2d_poincare}b). 
These Poincar\'{e} plots show a much more complicated pattern than similar plots evaluated for the paraboloidal basin (see for example figure $11$ from \citet{Maas2005my}). 
The appearance of more elaborate combinations of super and sub critical reflections is due to the presence of both convex and concave regions of the boundary, respectively the outer and the inner sphere \citep{Dintrans1999}.

\subsection{Three dimensional ray behaviour in shell geometries ($\phi_0\neq 0,\pi$)}
\label{sec:3d}
\begin{figure}
\centerline{
a)
 \includegraphics[width=.3\linewidth]{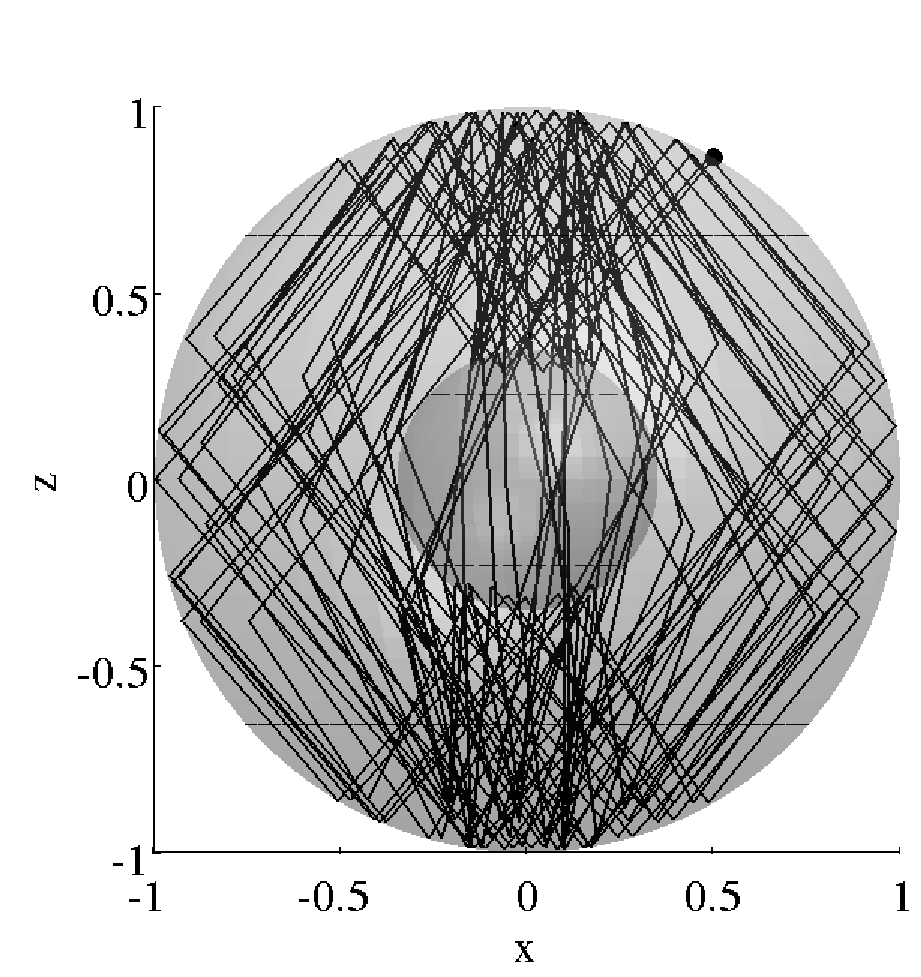}
b)
 \includegraphics[width=.3\linewidth]{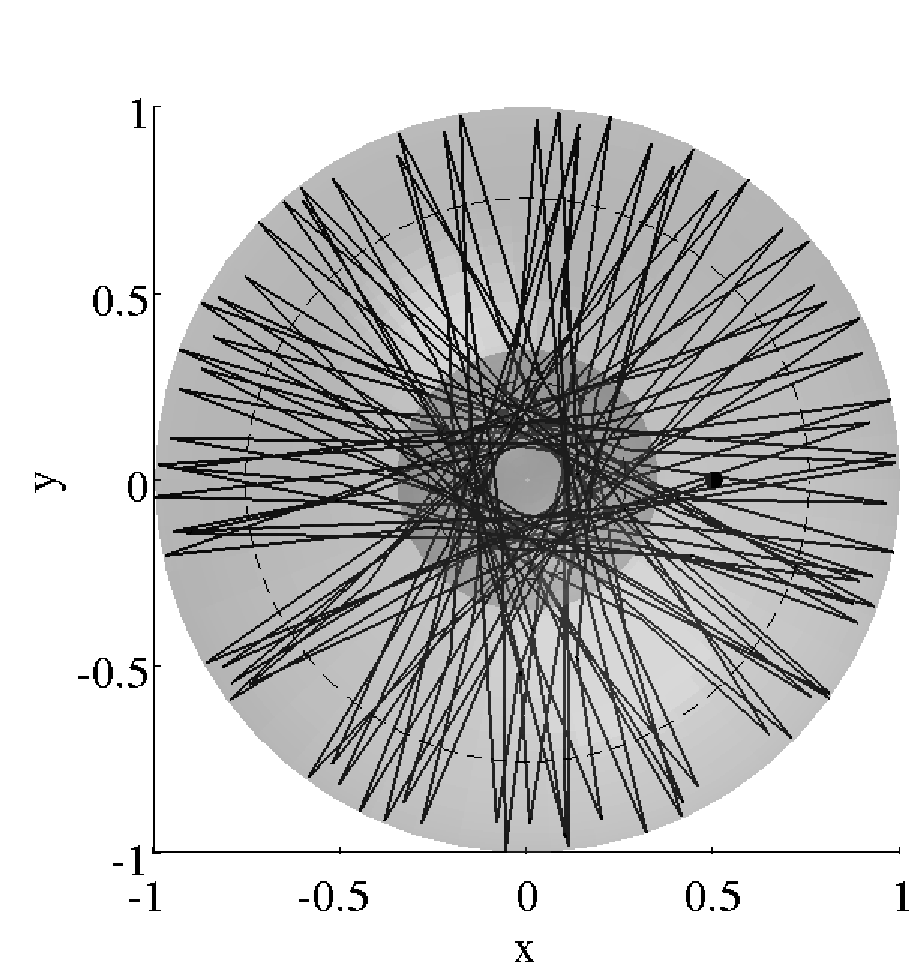}
c)
 \includegraphics[width=.3\linewidth]{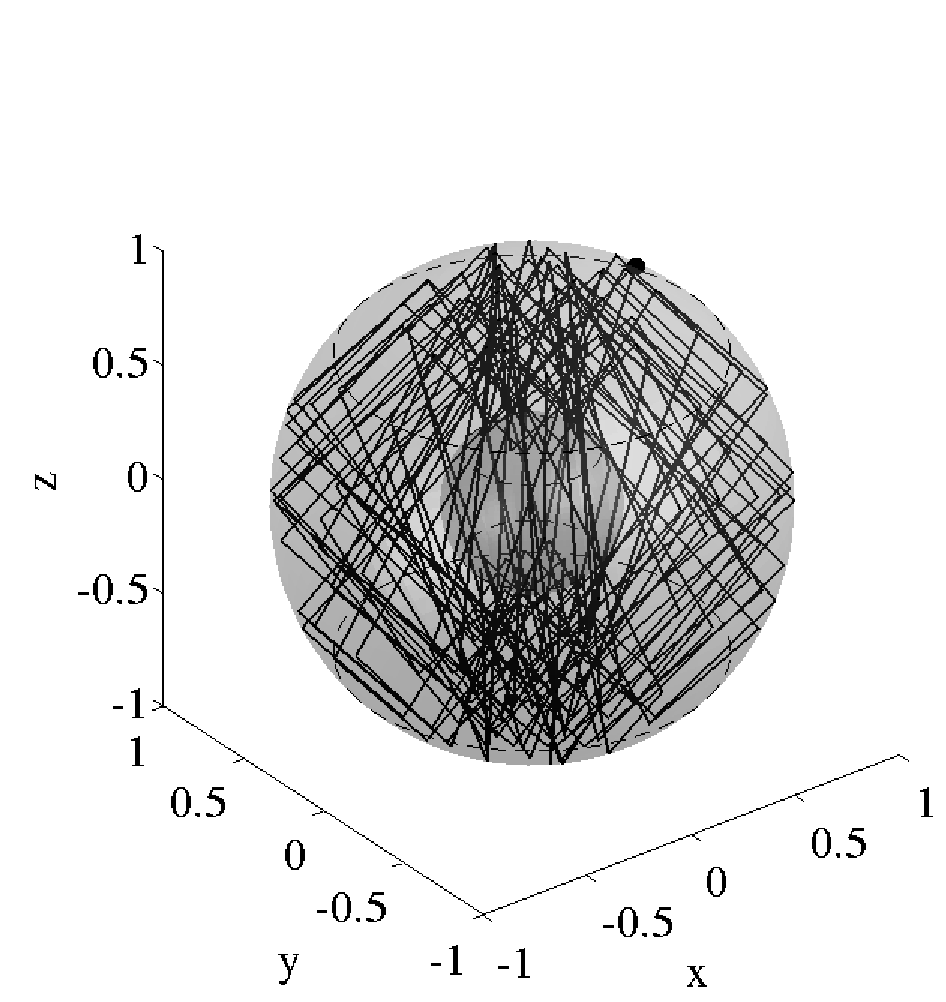}}
\caption{Example of an ergodic orbit in a three dimensional shell ($\eta = 0.35$). For $\omega = \sqrt{\frac{3}{7}}$ the ray is launched at the outer sphere at $x_0 = 0.15 $ , $\phi_0 = -7\pi/8$, and followed for 200 reflections. (a) meridional view ($x,z$ plane), (b) top view ($x,y$-plane) , (c) full 3D view ($x,y,z$ perspective). Black circles correspond to critical latitudes.  Black dot corresponds to the launching position $x_0$.\label{fig:3d_erg}}
\end{figure}
\begin{figure}
\centerline{
a)
 \includegraphics[width=.3\linewidth]{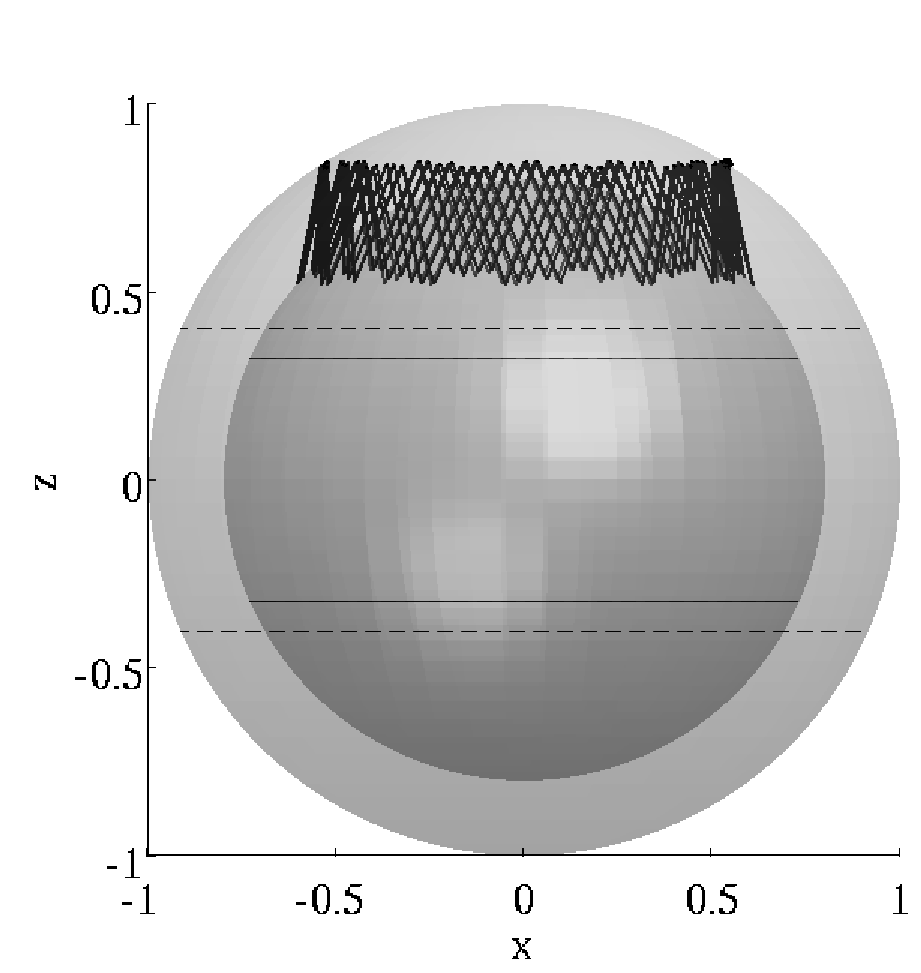}
b)
 \includegraphics[width=.3\linewidth]{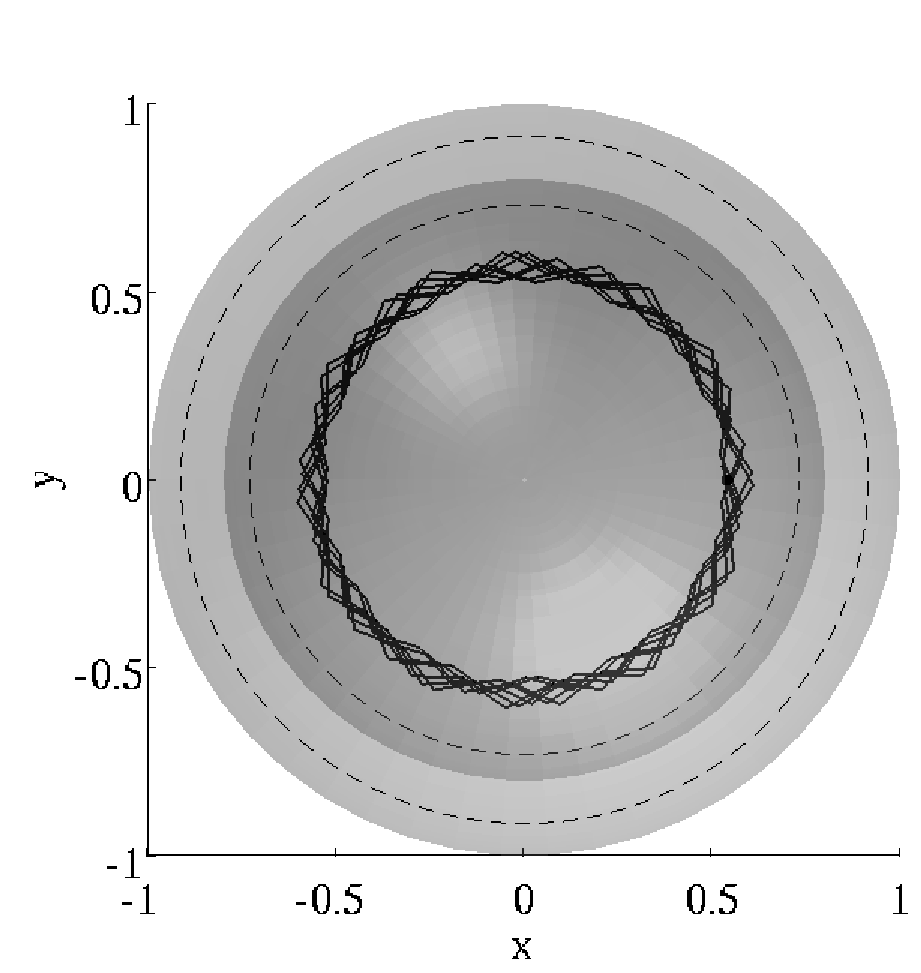}
c)
 \includegraphics[width=.3\linewidth]{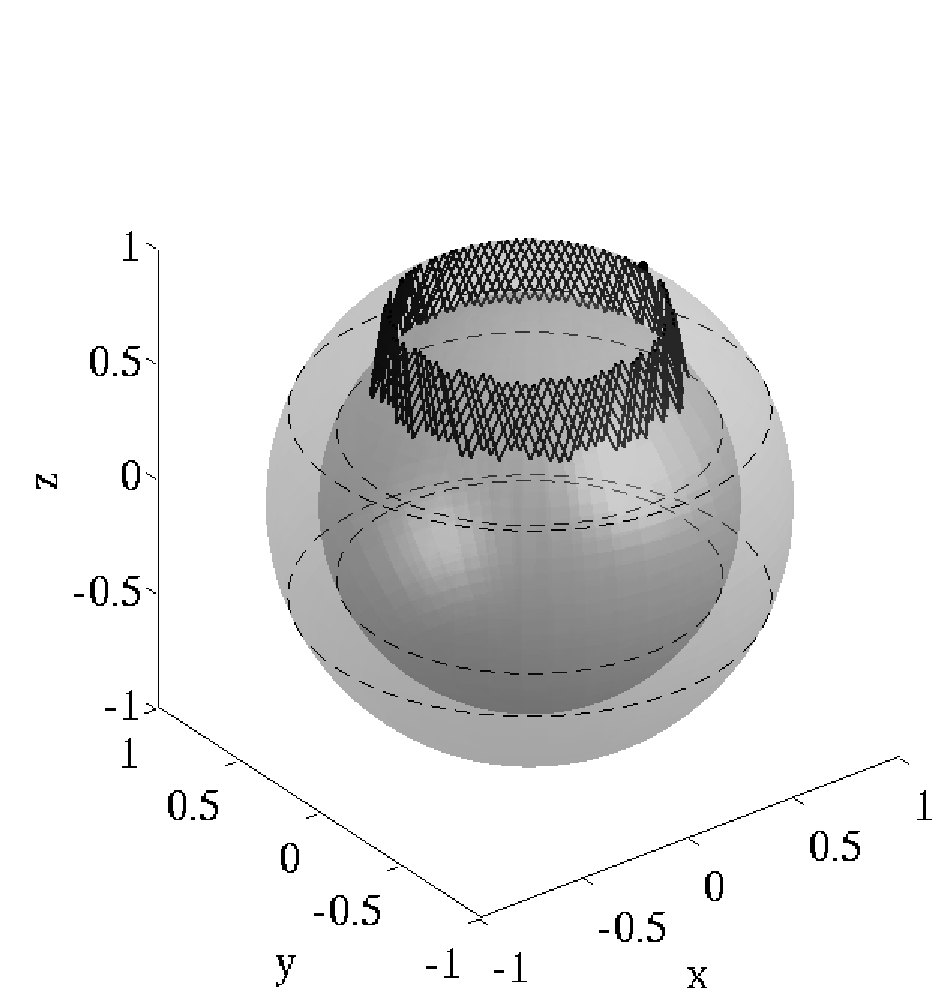}}
\caption{As figure \ref{fig:3d_erg} for a quasi-periodic, non converging trajectory. Here $\eta = 0.8$, $\omega = 0.4051$, $x_0 = 0.54$ , $\phi_0 = 1.342$, $200$ reflections. This trajectory corresponds to the star in figure \ref{fig:scan_xphi2}a. \label{fig:eta8white}}
\end{figure}

\begin{figure}
\centerline{
a)
 \includegraphics[width=.3\linewidth]{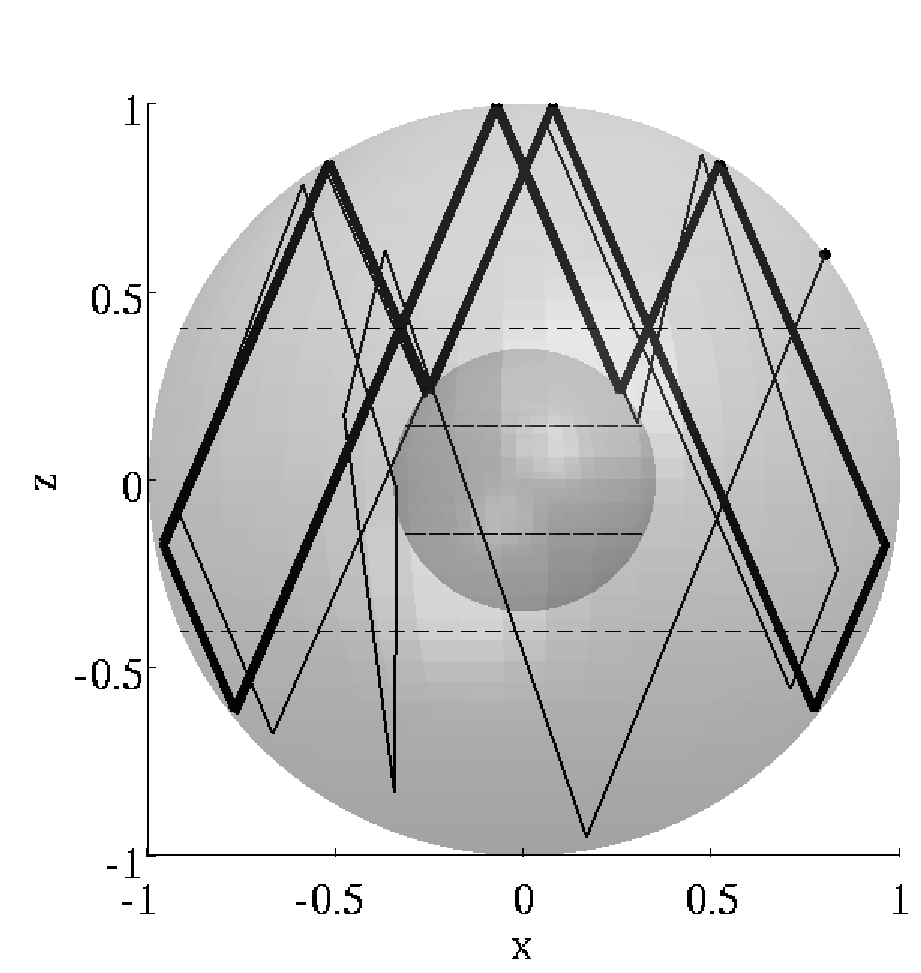}
b)
 \includegraphics[width=.3\linewidth]{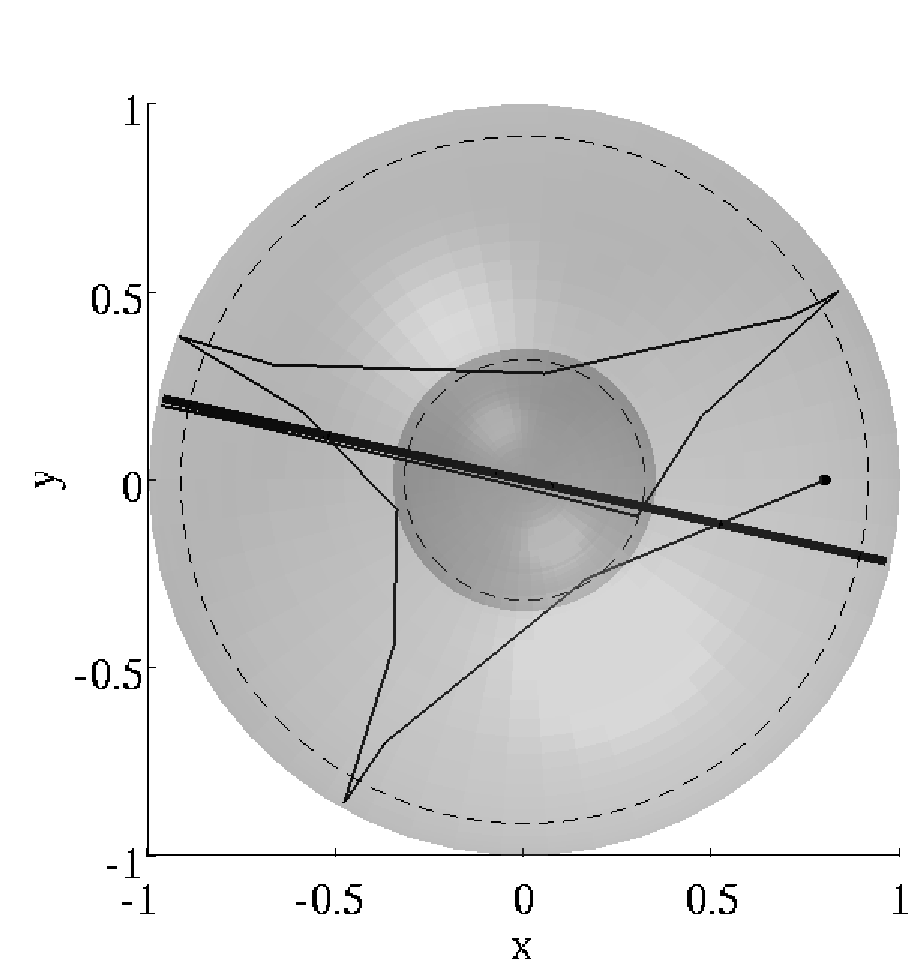}
c)
 \includegraphics[width=.3\linewidth]{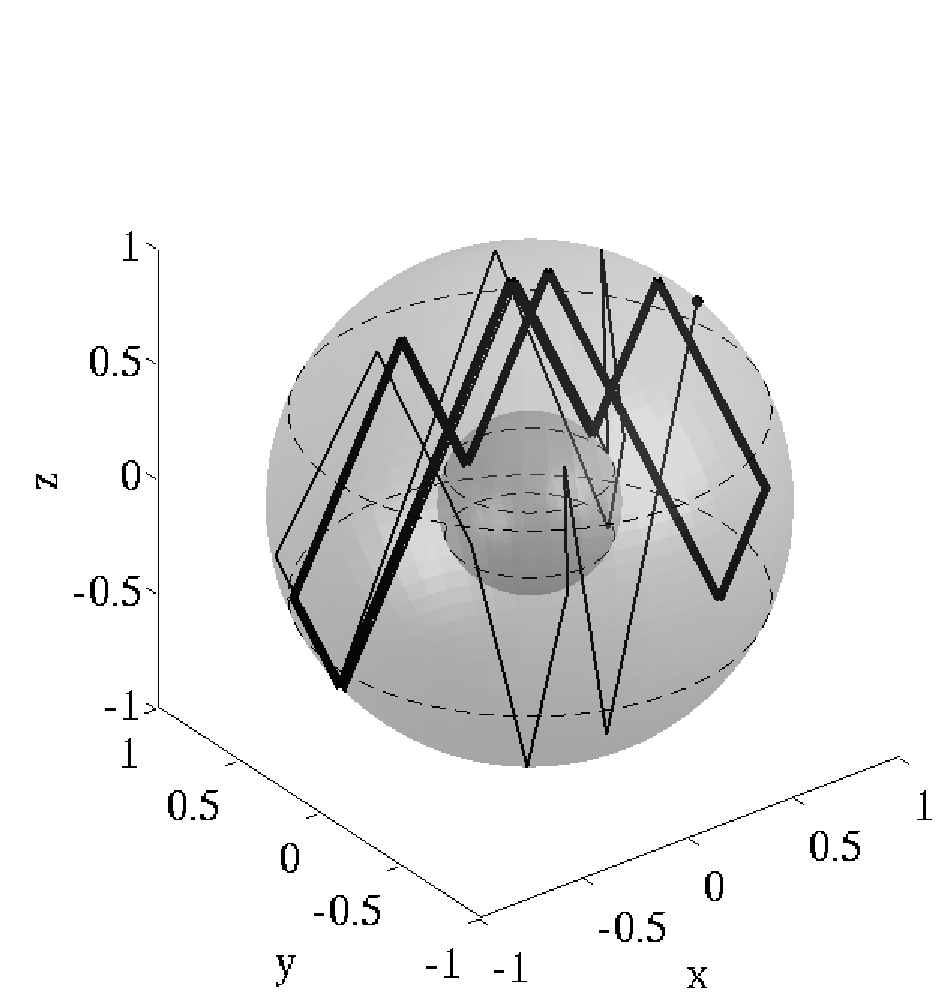}}
\caption{As figure \ref{fig:3d_erg}, but for a meridional attractor in the three dimensional shell ($\eta = 0.35$). Here the ray at $\omega = 0.4051$ is launched at $x_0 = 0.8$ , $\phi_0 = -7\pi/8$. Thick line marks the closed cycle of the final wave attractor. This trajectory also constitutes an example of a polar three-dimensional attractor (see text). \label{fig:3d_att}}
\end{figure}
In the following part of this section, differently from the approach that has been presented so far, ray motion will no longer be  constrained to a meridional plane, and results will be presented for ray tracing of inertial waves in a fully three dimensional spherical shell geometry. 

Doing so, we allow for azimuthal inhomogeneities to develop: waves, initially forced with a zonal propagation component, are subject to focusing and defocusing reflections from the boundaries and possibly refract towards a meridional plane, eventually becoming trapped in that plane. This mechanism, hypothesized by \citet{Maas2001} and \citet{Maas2007my}, is here observed for the first time.
Three dimensional trajectories can be interpreted as follows.
If a trajectory (fully determined by the domain geometry, launching position $\vec{x}_0$, launching direction $\phi_0$ and frequency $\omega$) is launched in a meridional plane (at $\vec{x}_0=0$, $\forall \phi_0$, or alternatively at $\phi_0 = 0, \pi$, $\forall \vec{x}_0$), it will never leave the plane, even if a full three-dimensional algorithm is used. This class of solutions corresponds to the known class of purely meridional trajectories.
On the other hand, if a trajectory is launched outside a meridional plane (all other combinations of $\vec{x}_0$ and $\phi_0$), either it will never cross the basin on a meridional plane (new class of zonally propagating solutions), or it will asymptotically approach one particular plane (occurrence of a meridional attractor), rendering the trajectory indistinguishable from a purely meridional trajectory after an appropriate number of reflections.

Results from three dimensional analysis are hard to be effectively presented on a two dimensional sheet of paper. In this work three perspectives for each example of trajectory will be presented: a meridional view (usually the $x,z$-plane), the top view ($x,y$-plane) and a full 3D perspective view ($x,y,z$). Note that the apparent change in vertical orientation of a wave ray is a visual effect due to the projection only; the ray always obeys equation (\ref{eq:dispersion2}), but in a three dimensional fashion.
In analogy with the two dimensional studies, rays have different behaviour according to their launching position $\vec{x}_0$, launching direction $\phi_0$, frequency $\omega$ and width of the spherical gap. Moreover, the combination of $\vec{x}_0$ and $\phi_0$ influences the horizontal final orientation of the possibly occurring meridional attracting plane.
In the following, we will refer to this final horizontal orientation of the meridional attractor as to $\phi_\infty$, an angle measured anticlockwise with respect to the $x$-axis, whose arbitrary orientation is defined by the location of the initial launching point ($x_0,0,z_0$).

Three kinds of behaviour are observed: the orbit can be domain filling (ergodic-like, figure \ref{fig:3d_erg}); it can be quasi-periodic, its path filling a portion of the domain only (regular pattern, figure \ref{fig:eta8white}), or it can eventually be trapped, first onto a meridional plane (meridional attractor), and subsequently, within that plane, on a two-dimensional limit cycle (attractor, figure \ref{fig:3d_att}). 
It is worth noticing that here only trajectories reflecting from both inner and outer boundaries are listed. Of course, for certain combinations of $\eta,\omega,\vec{x}_0,\phi_0$, rays exist that do not touch the inner sphere at all. These trajectories can be thought of as living in a full sphere ($\eta=0$), and we will discuss them separately in \S\ref{sec:edge} and \ref{sec:sphere}.
Differently from the two-dimensional case, no three-dimensional periodic orbit has been observed so far. 
This is probably due to the appearance in the three-dimensional framework of an extra parameter in the problem: the initial launching direction. Whereas in the two-dimensional problem the only parameters are $\eta$, $\omega$ and $\vec{x}_0$, and periodicity is determined by $\omega$ only ($\phi_i$, at every $\vec{x}_i$, being either $0$ or $\pi$). Surprisingly or not, in the three-dimensional case the combination of $\vec{x}_0$ and $\phi_0$ (and therefore subsequent pairs $\vec{x}_n,\phi_n$) influences the occurrence of meridional trapping (and, in case, the final orientation $\phi_\infty$) and plays a crucial role in preventing/allowing a trajectory to close exactly onto itself.
 No universally valid relation between $\omega$ and $\phi_0,\vec{x}_0$ has been found so far to compute three-dimensional periodic orbits in a spherical shell, but we cannot exclude their presence.
Obviously, because of the rotational symmetry of the problem and of the arbitrarily positioned $x$-axis, if one combination of $\omega, x_0, \phi_0$ exists for which the trajectory is closed, an infinite number of closed trajectories will exist in the same domain.

In analogy with the two-dimensional case, three-dimensional ray tracing thus allows us to explore singular solutions occurring in the shell, but does not say anything about the possible existence of regular modes in the domain.
The occurrence of meridional attractors in the inviscid model strengthens the validity of all previous studies on inertial wave attractors in geophysical and astrophysical frameworks \citep{Bretherton1964, Friedlander1982a, Dintrans1999, Rieutord2001, Maas2007my}.
The focussing power of an attractor is not limited to its two dimensional plane, but can act in some geometries, such as the shell, in a three dimensional fashion as well.
The appearance of attractors in a spherical shell resembles what has been shown for a paraboloidal basin by \citet{Maas2005my} with an analogous ray tracing study. The latter study has found confirmation in laboratory experiments \citep{Hazewinkel2010a}, and, for a parabolic channel geometry, in a (viscous) numerical experiment \citep{Drijfhout2007}.
All these works corroborate the existence and the power of three dimensional attractors, and suggest they could play an important role in non regular geometries (symmetry breaking geometries, as opposite to spherical or ``flat'' rectangular geometries), in focussing energy onto specific and predictable locations, triggering crucial mixing phenomena in all sorts of stratified, rotating fluids \citep{Swart2010}.

As can already be noticed from figure \ref{fig:3d_att}, the number of reflections needed for the attractor to take place is relatively high, and the chances to see even one or two reflections in realistic media are quite low. Apart from observations in laboratories, there are no observations in Nature, so far, except for the ubiquitously observed peak at the local Coriolis frequency ($2\Omega\sin \phi$, where $\phi$ is now the latitude of the measurements) which bears evidence, in the stratified case, of a point attractor \citep{Maas2001, Gerkema2005b}.
Nevertheless, as will be shown in the following section, the trapped energy is collected over a broad range of possible latitudinal input locations and initial launching directions of the perturbation, and this makes unnecessary for the attractor to be fully developed in order to have a significant increase of energy in a restricted longitudinal range (see corresponding experimental evidences of complicated attractors in \citet{Hazewinkel2010a}).   

\subsection{Meridional attractors}
\label{sec:attractor}
\begin{figure}
\centerline{
a)
 \includegraphics[width=.3\linewidth]{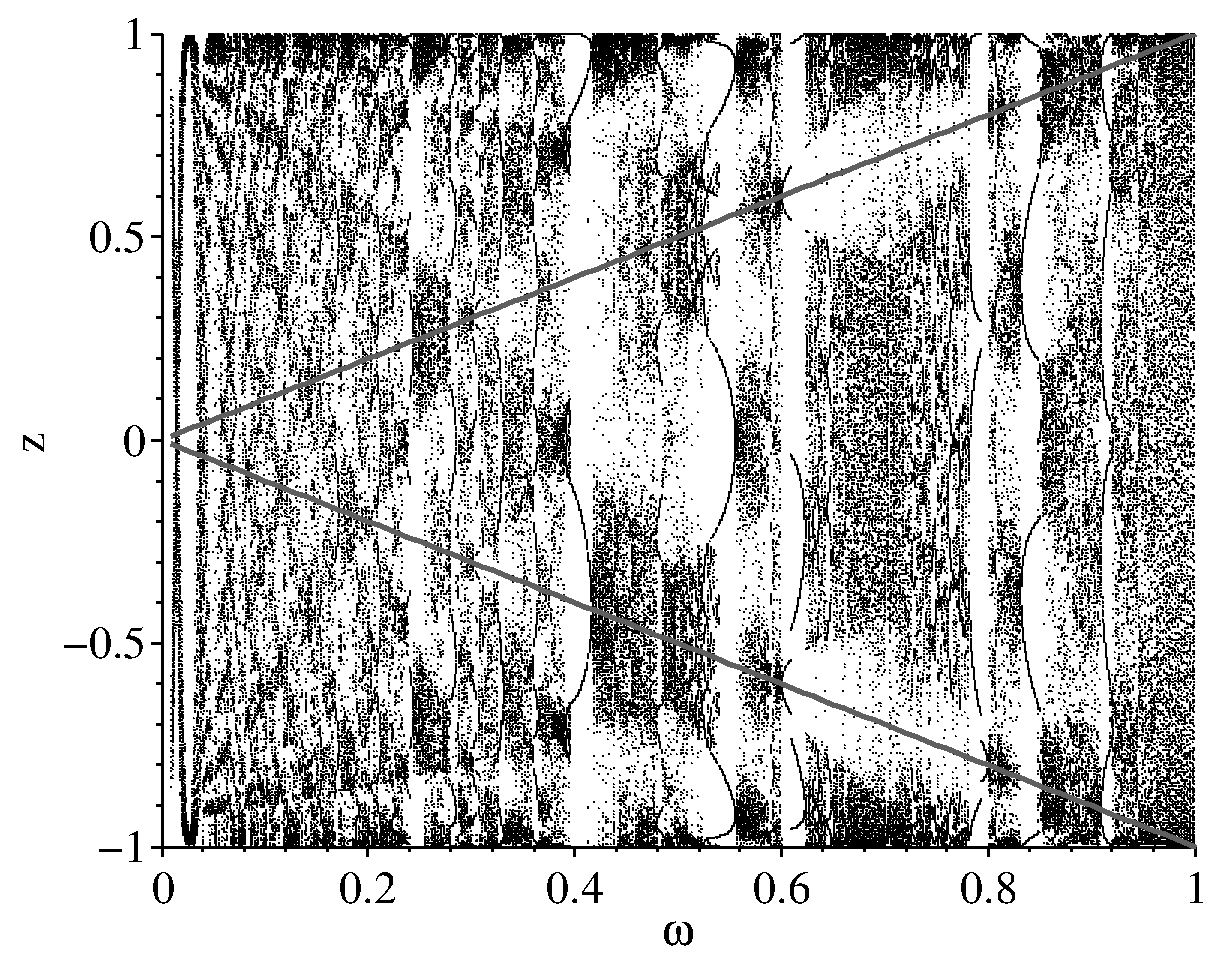}
b)
 \includegraphics[width=.3\linewidth]{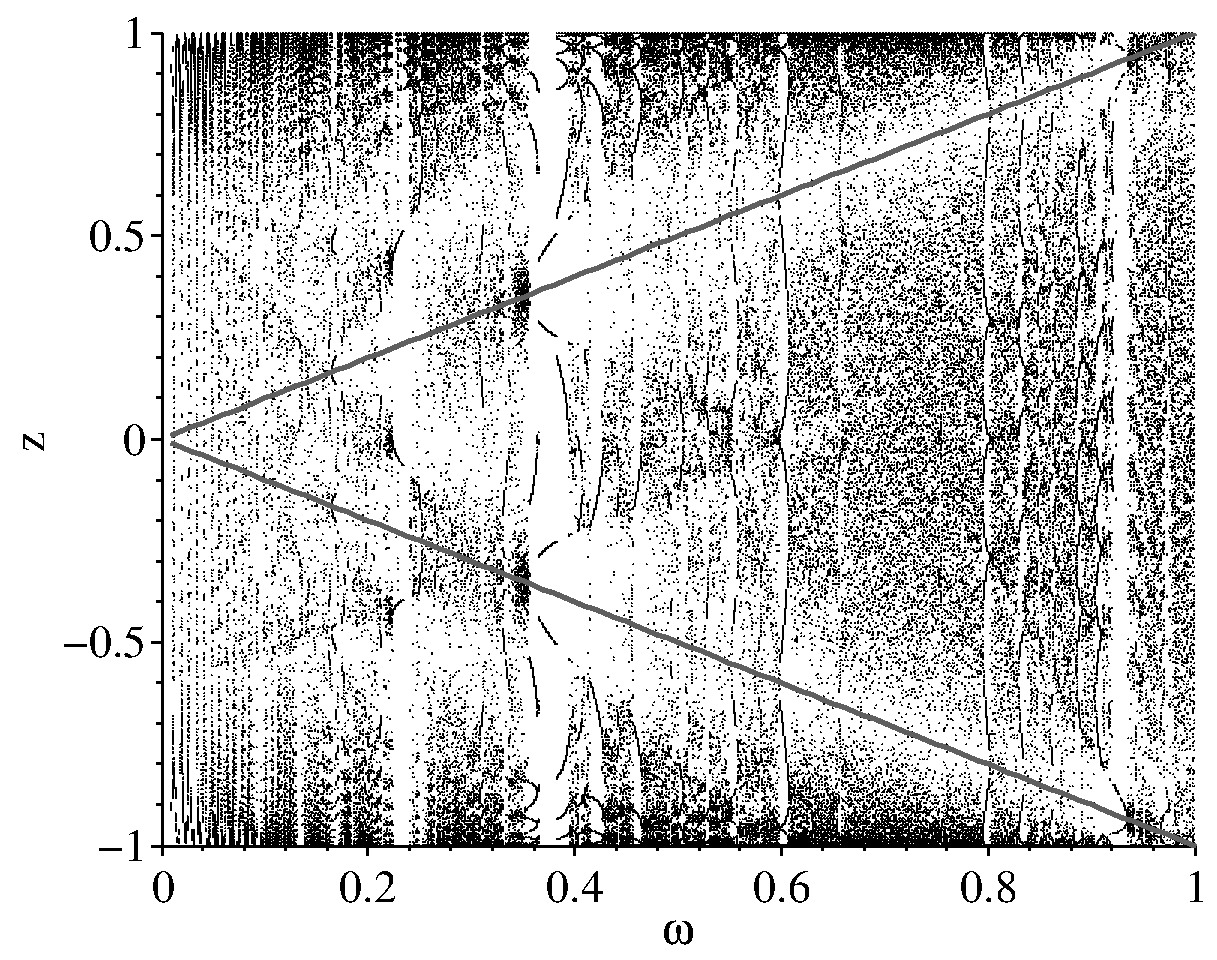}
c)
 \includegraphics[width=.3\linewidth]{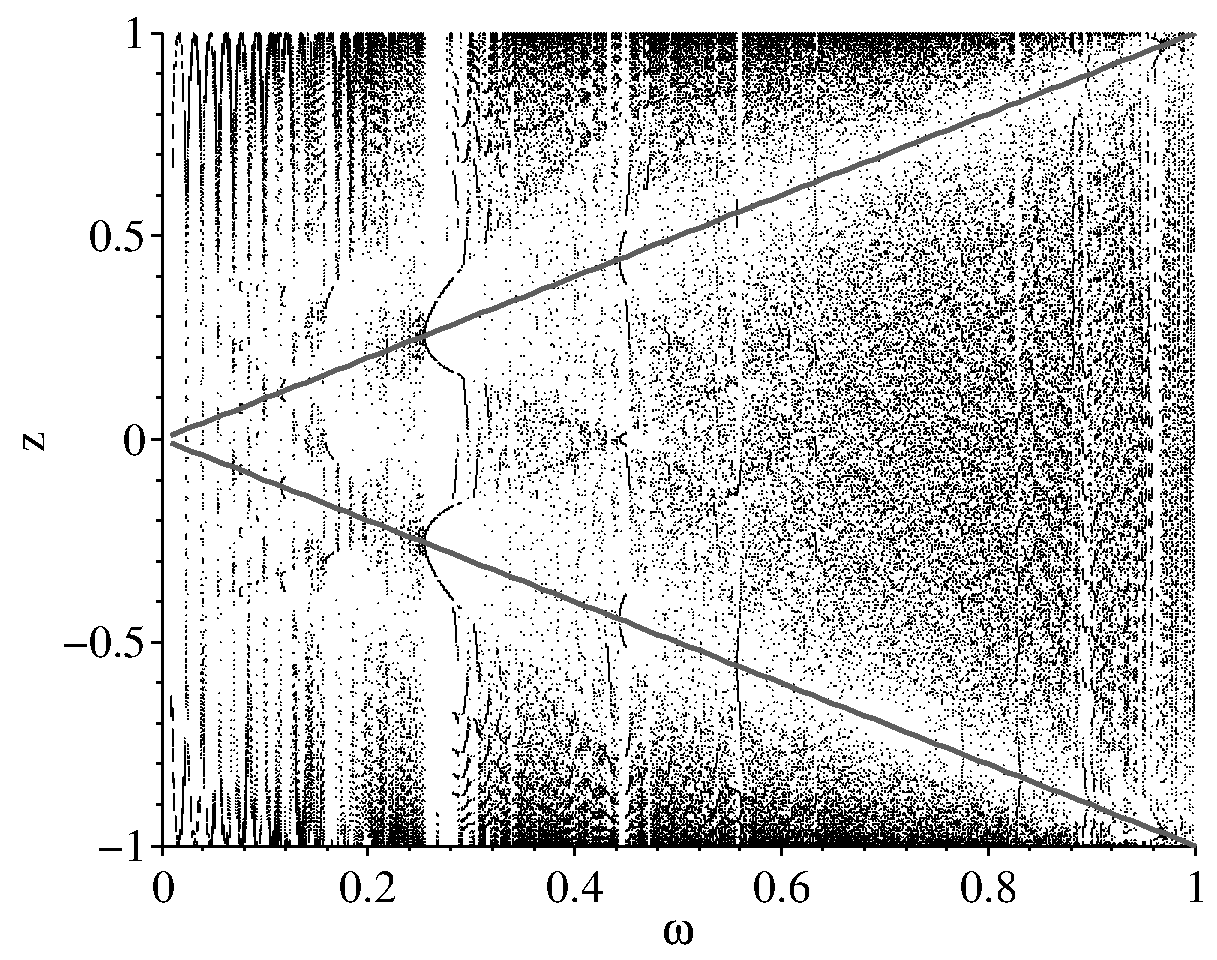}}
\caption{Poincar\'{e} plots for shell thickness $\eta = 0.35$ (a), $0.8$ (b) and $0.9$ (c). Here $x_0=0.2$ on the outer sphere, $\phi_0 = 5\pi/4$. On the $x-$axis, $\omega$ in the inertial range. On the $y-$axis, the $z$ coordinate of reflection points on the outer sphere (``surface'') of the last $20$ reflections (out of $1000$), both hemispheres. Grey lines represent the locations of the critical latitude for each $\omega$. \label{fig:3d_poincare}}
\end{figure}
% XPHI 35 
\begin{figure}
\centerline{
a)
 \includegraphics[width=.5\linewidth]{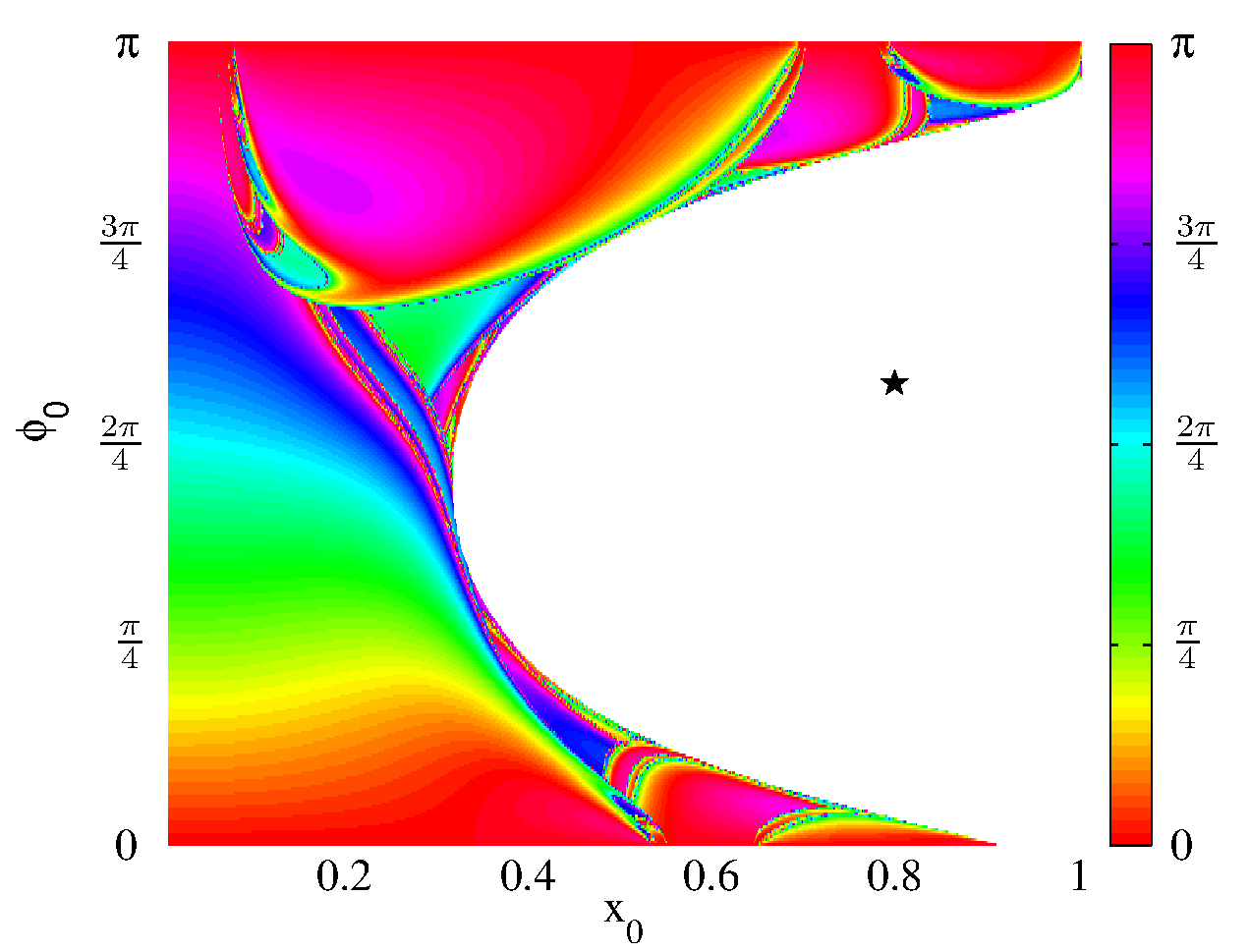}
b)
 \includegraphics[width=.5\linewidth]{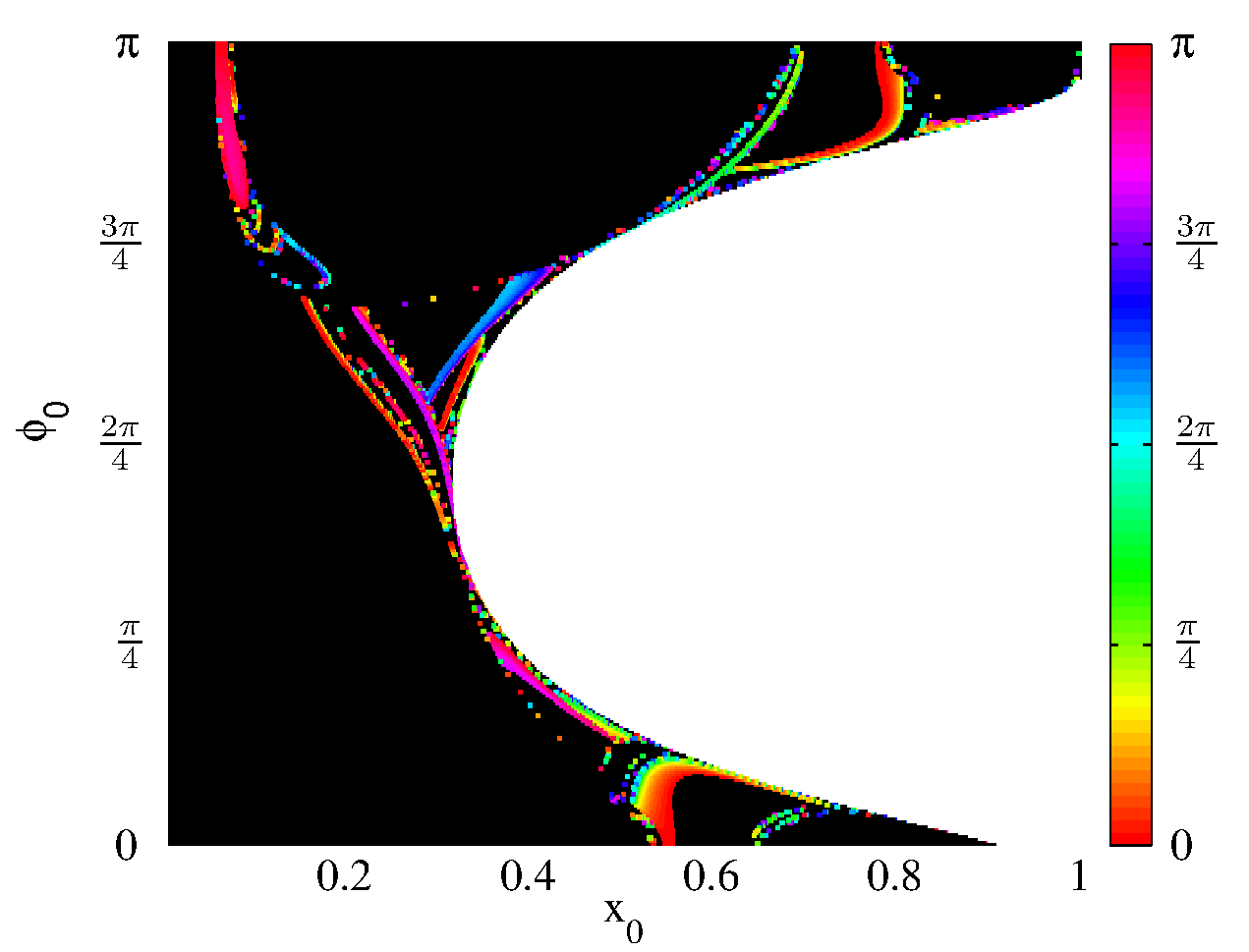}}
\caption{(a) plot of $\phi_\infty$ (colour, see legend in radians) as function of launching position and direction $x_0, \phi_0$, for $\eta = 0.35$, after $1000$ reflection. The perturbation frequency $\omega$ is taken to be equal to $0.4051$ (see white vertical band in figure \ref{fig:3d_poincare}). On the $x$-axis, all possible $x_0$ are scanned, from zero to one. On the $y$-axis, $\phi_0$ is scanned just between $0$ and $\pi$, for symmetry reasons. Colours correspond to meridional trapping, white to zonally propagating rays. Black star in figure \ref{fig:scan_xphi}a corresponds to trajectory displayed in figure \ref{fig:eta35white}. (b) Same as figure (a) where black corresponds to coloured area in (a), hence to meridional trapping, whereas colours now represent \textit{equatorial} type of attractors only (see \S\ref{sec:eqtVSpol} for details), and their final horizontal orientation $\phi_\infty$. An example of \textit{equatorial} attractor is visible in figure \ref{fig:eqt3d_att}. \label{fig:scan_xphi}}
\end{figure}
% XPHI HIGHER ETA
\begin{figure}
\centerline{
a)
 \includegraphics[width=.5\linewidth]{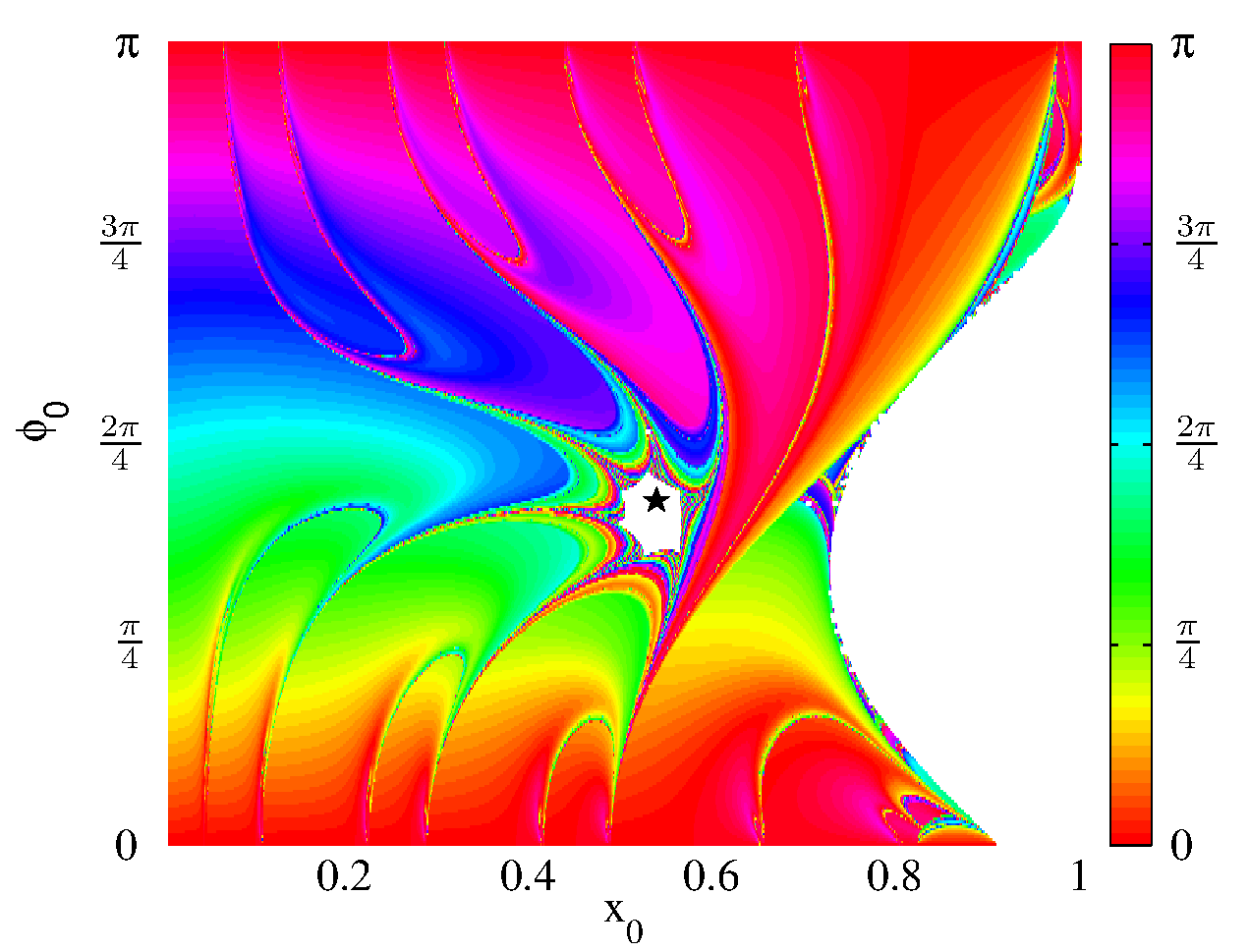}
b)
 \includegraphics[width=.5\linewidth]{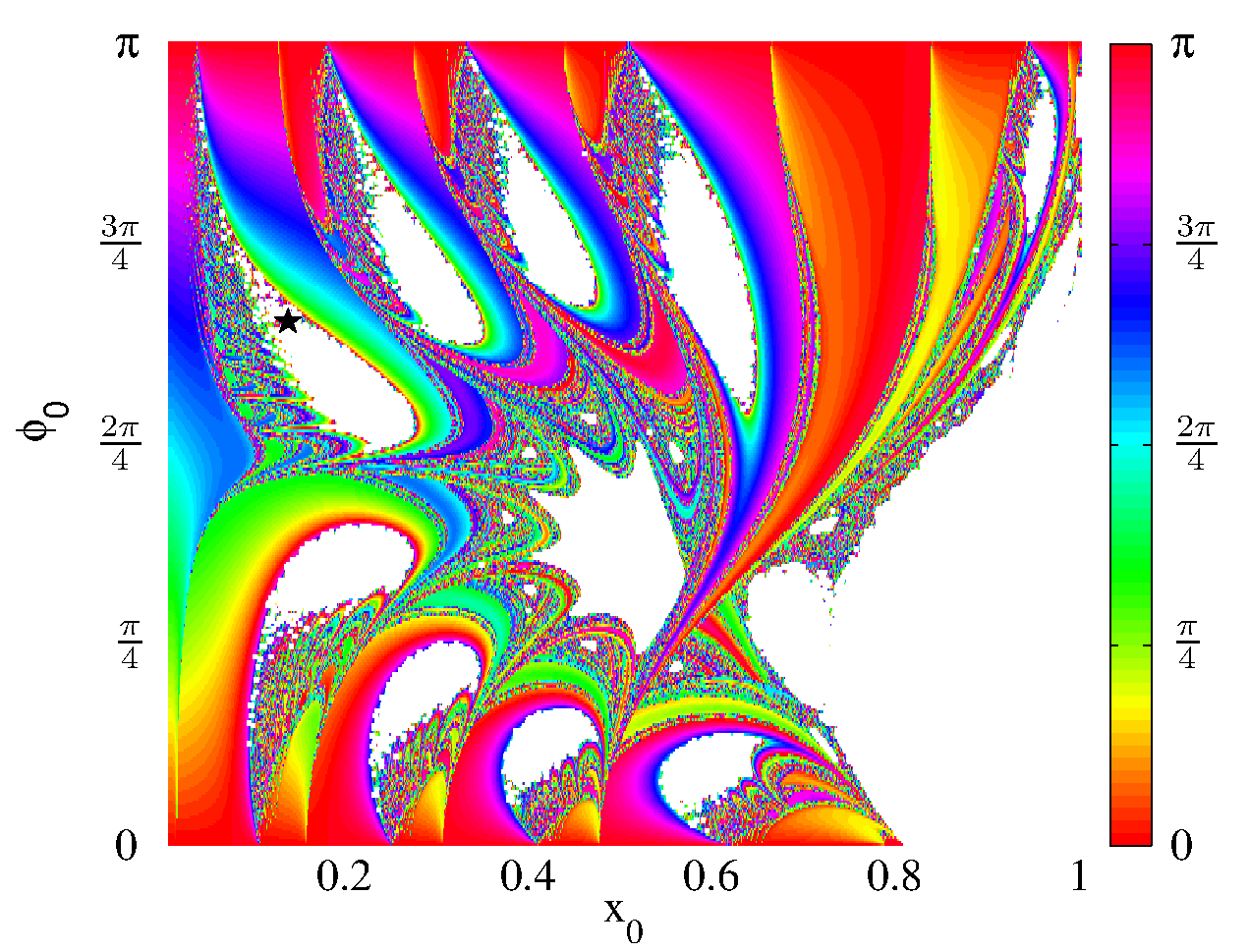}}
\caption{Same as in figure \ref{fig:scan_xphi} for $\eta = 0.8$, $\omega= 0.4051$ (a) and $\eta = 0.9$, $\omega= 0.5842$ (b), after $5000$ reflections. On the $x$-axis, all possible $x_0$ are scanned, from zero to one.  On the $y$-axis, $\phi_0$ is scanned just between $0$ and $\pi$, for symmetry reasons. Colours represent meridional trapping, white areas non converging regions. Black star in figure \ref{fig:scan_xphi2}a  corresponds to trajectory displayed in figure \ref{fig:eta8white}. Black star in figure \ref{fig:scan_xphi2}b  corresponds to trajectory displayed in figure \ref{fig:eta9white}.\label{fig:scan_xphi2}}
\end{figure}

In this section we will explore the parameter space for the shell case study, in order to show that meridional attracting planes (1) are not exceptional, (2) have $\eta$ (geometry), $\omega$ (frequency), $x_0$ and $\phi_0$ (initial conditions) dependencies, and (3) occur after a varying number of reflections (due to different focussing power of different combination of  $\eta$, $\omega$, $x_0$ and $\phi_0$). 

It is worth stressing that the occurrence of meridional attractors constitutes only the first step of the focussing process possibly experienced by inertial waves. Once the ray motion is restricted to a meridional plane, ``classical'' two-dimensional attractors arise, confining wave energy to their limit cycles within that plane, from which they can no longer escape.
These ``classical'' two dimensional attractors have been broadly explored for the spherical shell case since \citet{Stewartson1972}, in both homogeneous and stratified rotating fluids \citep{Rieutord2001, Maas2001}. Therefore they will not be subject of the current analysis.

The appearance of meridional attractors is evident from the Poincar\'{e} plots (figure \ref{fig:3d_poincare} for three different $\eta$ values) resulting after a three-dimensional ray tracing in the shell. In these Poincar\'{e} plots, on the horizontal axis the inertial frequency range is scanned ($0<\omega<1$). On the vertical axes, the $z$ coordinates of the reflection points on the outer sphere (``surface'') of the last $20$ reflections (out of $100,1000,1000$ reflections respectively) are depicted. Grey lines represent locations of the corresponding critical latitude for each value of $\omega$. 
Different frequency windows representing ``simple'' attractors (characterized by a small number of boundary reflections) emerge for different values of shell thickness, visible as white vertical bands.
We observe that even for the three dimensional meridional attractors the critical latitudes act as repellors for the rays, and their repelling power seems to increase with $\eta$. 
Note that figures \ref{fig:3d_poincare}a and \ref{fig:2d_poincare}b are similar, but not identical, meaning that the frequency windows of the attractors are the same in the two- and in the three-dimensional cases, but the final attractor trajectories can show a different projection on the $\hat{z}$-axis, according to the final orientation of the attracting plane.

In figures \ref{fig:scan_xphi}a (for $\eta = 0.35$),  \ref{fig:scan_xphi2}a (for $\eta = 0.8$) and \ref{fig:scan_xphi2}b (for $\eta = 0.9$) the fate of a whole characteristic cone is depicted after respectively $1000$, $5000$ and $5000$ reflections, for frequencies for which meridional attractors occur according to figure \ref{fig:3d_poincare}.
On the horizontal axis of these figures, all possible $x_0$ are considered, from zero to one. The perturbation is always launched on the surface of the outer sphere, in the northern hemisphere, $y_0 = 0$ and $z_0 = \sqrt{1 - x_{0}^{2}}$, with an initial negative vertical velocity.
On the vertical axes of figures \ref{fig:scan_xphi}a, \ref{fig:scan_xphi2}a and \ref{fig:scan_xphi2}b, for symmetry reasons, $\phi_0$ is considered between $0$ and $\pi$ only.
Colours indicate the presence of meridional trapping, and represent orientation of the final attracting plane ($\phi_\infty$), following the colours legend. White areas correspond to trajectories that are not subject to meridional focussing and they are interpreted as zonally propagating waves (see \S\ref{sec:edge} for comments).  
A single vertical column in figures \ref{fig:scan_xphi}a, \ref{fig:scan_xphi2}a and \ref{fig:scan_xphi2}b can be read as the final longitudinal location of the energy for the whole three dimensional cone excited at a single point on the surface: the three figures present all, for small $x_0$ (``polar'' source), a horizontally striped structure, meaning that the meridional attractor of a single three-dimensional ray will preferentially approach a plane that has the same horizontal orientation as the original launching direction ($\phi_\infty \approx \phi_0$). 
This, in principle, would lead to a zonally homogeneous distribution of energy in the domain.
Remarkably, the more equatorward we move the source, the more stripes are deformed and one final orientation of the attractors prevails (see for example the range $x_0 = [0.7-0.8]$ for figure \ref{fig:scan_xphi2}b, for which $\phi_\infty \sim \pi$ for $\phi_0\gtrsim\pi/2$ and no attractor occurs for $\phi_0<\pi/2$). 
This tells us that, from the whole excited cone, containing rays with initial directions $\phi_0 \in [-\pi, \pi]$, most of the rays are meridionally trapped on planes that will have approximately the same orientation in the $x,y$-plane. Higher energy values are therefore expected in correspondence with those specific longitudinal ranges.
This result is supported by observations performed in a three dimensional paraboloidal basin by \citet{Hazewinkel2010a}. In this experiment, a stable uniform stratification in density was disturbed by an off-centred oscillating sphere, and internal gravity waves were excited in a paraboloidal basin. A three-dimensional tomographic reconstruction of the amplitude (energy) distribution of these waves in the basin has shown the occurrence of a preferential vertical trapping plane.
A centred wave source in the paraboloid would produce no preferential vertical plane, since energy would spread equally along the longitudinal coordinates. 
This is true for the shell as well, since a centred (``polar'') launching position of the ray will result in a purely meridional ray motion, that doesn't allow for longitudinal inhomogeneities.   

We conclude that the present three-dimensional ray tracing study therefore not only confirms what was already shown by \citet{Bretherton1964} and \citet{Stewartson1971} and \citet{Stewartson1972}, that is that singular solutions (corresponding to attracting orbits) characterize the internal wave field in a shell domain. The present study extends the validity of this finding to cases in which a local source is considered, and three dimensional effects are taken into account.

Computing three-dimensional orbits clearly shows how, decreasing the thickness of the shell, a larger number of reflections is needed in order to completely develop attractors in the domain. 
This can be interpreted as follows: the attracting plane emerges after a combination of supercritical and subcritical reflections of the ray throughout the domain. If the thickness of the shell is small, the ray has to experience several reflections to reach a supercritical (subcritical) region of the domain, because the possible path between two subsequent reflections is small compared to the extent of these regions. Conversely, if the shell is thick, the path between two subsequent reflections is large, and less reflections are needed for the ray to experience both focusing and defocusing reflections.
As already mentioned, this behaviour undermines the attracting power of attractors in thin shells (as the ocean on an aqua planet) and in more realistic (viscous, inhomogeneous) settings; nevertheless energy enhancement is possibly detectable even in absence of a fully developed attractor.
Other computational experiments have shown that $\phi_0$ also affects the number of reflections needed for the attractor to develop. This is because the amount of focusing depends on the incident angle.

\subsection{Equatorial and polar attractors}
\label{sec:eqtVSpol}
\begin{figure}
\centerline{
a)
 \includegraphics[width=.3\linewidth]{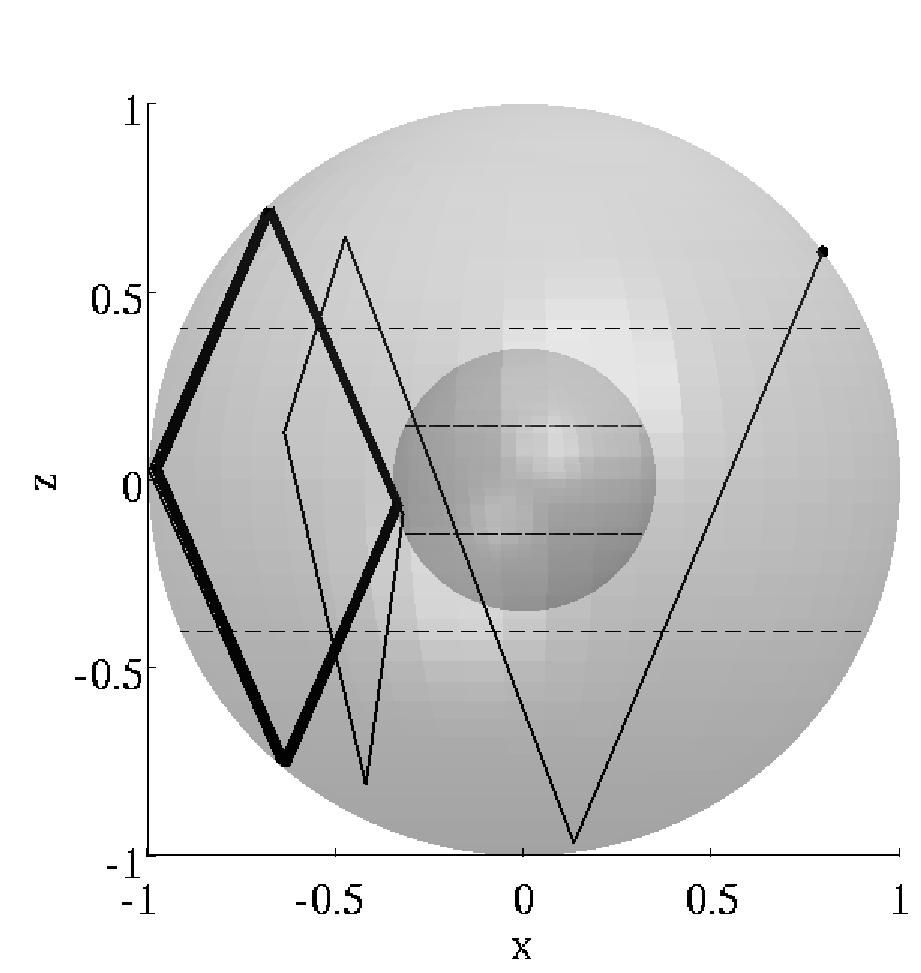}
b)
 \includegraphics[width=.3\linewidth]{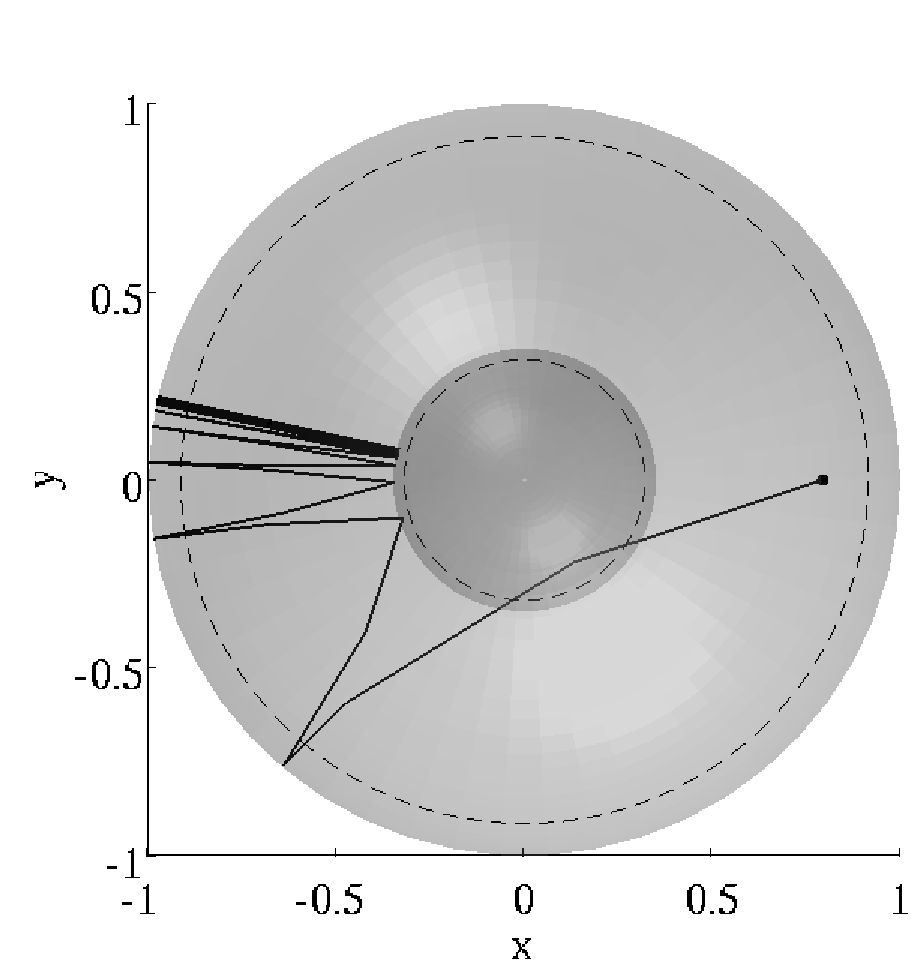}
c)
 \includegraphics[width=.3\linewidth]{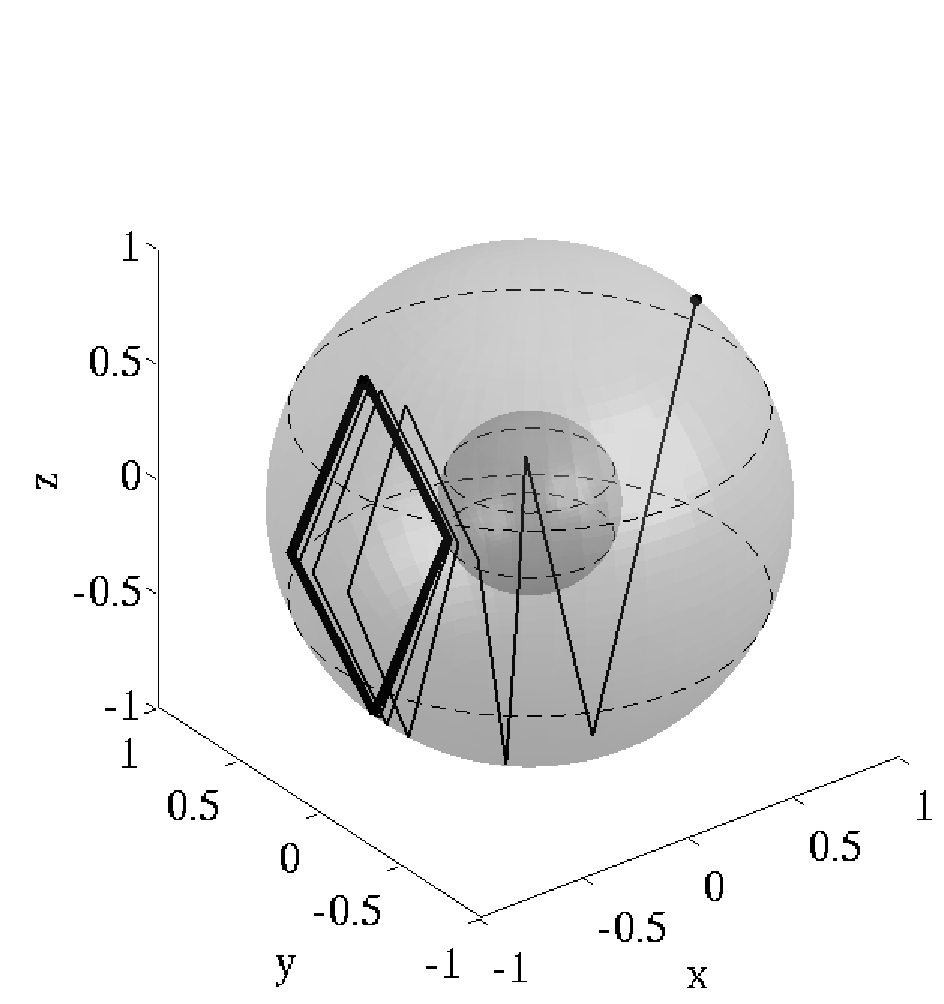}}
\caption{As figure \ref{fig:3d_erg}, but for $\eta = 0.35$, $\omega = 0.4051$, $x_0 = 0.794$ , $\phi_0 =  -2.822$, showing an equatorial three-dimensional attractor. The wave attractor is drown with a thicker line. \label{fig:eqt3d_att}}
\end{figure}
Fig. \ref{fig:3d_poincare} shows that singular meridional attractors occur in specific and predictable frequency bands, when the width of the shell is given. 
As we have mentioned before, the asymptotic shape of trapped, initially three-dimensional trajectories, are two-dimensional objects, whose structure exactly corresponds to the ones found by a simpler two-dimensional (meridional) ray study. Therefore, in analogy with the discussion by \citet{Rieutord2001}, we observe ``simple'' attractors (so called \textit{equatorial} attractors, in figure \ref{fig:eqt3d_att}), for example in the frequency band around $\omega=0.4$ in figure \ref{fig:3d_poincare}a, characterized by only four reflections with the boundary), coexisting with more ``complicated'' attractors (so called \textit{polar} attractors, in figure \ref{fig:3d_att}), characterized by a larger number of reflections.

Equatorial attractors occur in the low latitude range, and take place when reflections on the inner sphere occur below the inner critical latitude (supercritical reflections), whereas polar attractors, spanning from equatorial to polar regions, take place when reflections on the inner sphere occur poleward of the inner critical latitude (subcritical reflections).
Equatorial attractors appear more robust and energetically relevant than polar ones, because of the fewer reflections needed to build their closed cycle, but how often do they occur? Their occurrence surely depends on shell thickness, being more likely to hit inner supercritical regions if the inner sphere is larger. Moreover, additional radial density stratification increases their frequency band width \citep{Maas2007my}.
An example of the occurrence of equatorial attractors is shown in figure \ref{fig:scan_xphi}b, for the case $\eta=0.35$: black represents general meridional trapping (and corresponds exactly to the variously coloured areas in figure \ref{fig:scan_xphi}a); colours in figure \ref{fig:scan_xphi}b represent instead equatorial type of attractors only, and their horizontal orientation $\phi_\infty$, according to the colour map on the right hand side. As is clear by comparing figures \ref{fig:scan_xphi}a and b, equatorial attractors constitute the borders of the smooth areas in the parameter space. 

Because of their simple shape, equatorial attractors are good candidates to have a strong influence on low-latitude dynamics. Moreover, oceanic equatorial regions are among the longitudinally widest ocean basins, therefore energy contribution to this kind of structures could be collected in principle from a wide range of possible longitudinal atmospheric or tidal inputs, and is potentially of relevance to the still largely unexplained equatorial dynamics. 
 
\subsection{Zonally propagating waves}
\label{sec:edge}
\begin{figure}
\centerline{
a)
 \includegraphics[width=.3\linewidth]{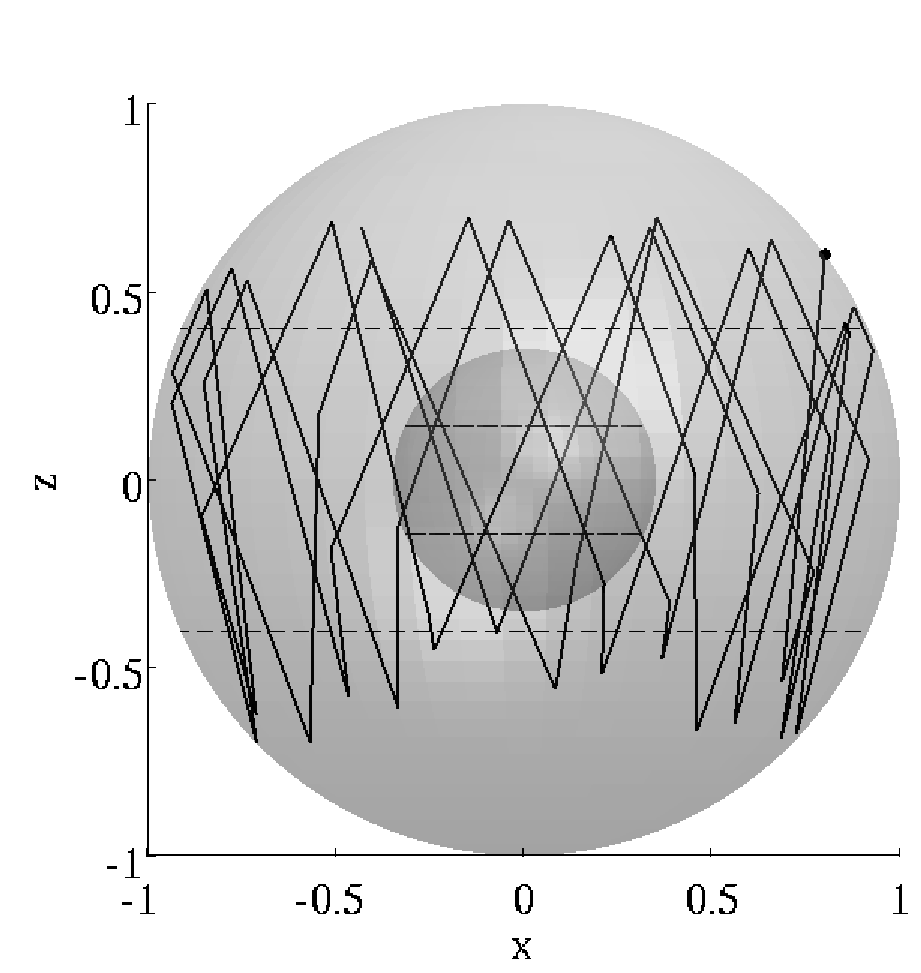}
b)
 \includegraphics[width=.3\linewidth]{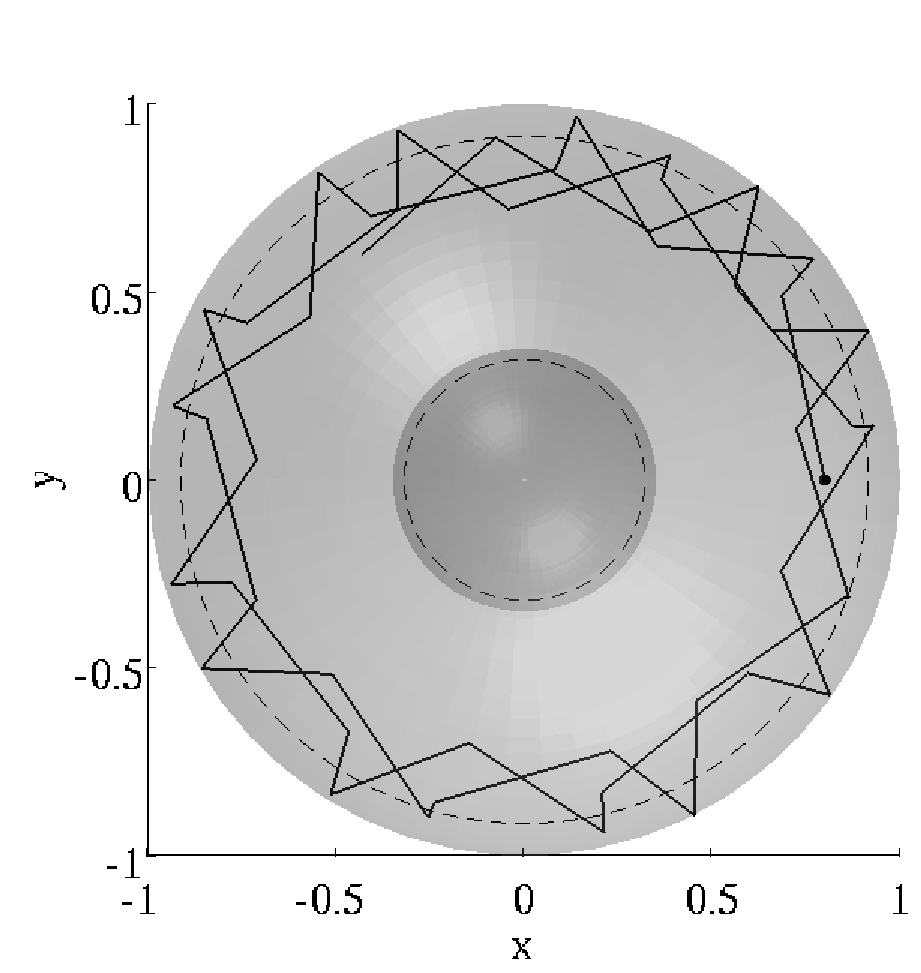}
c)
 \includegraphics[width=.3\linewidth]{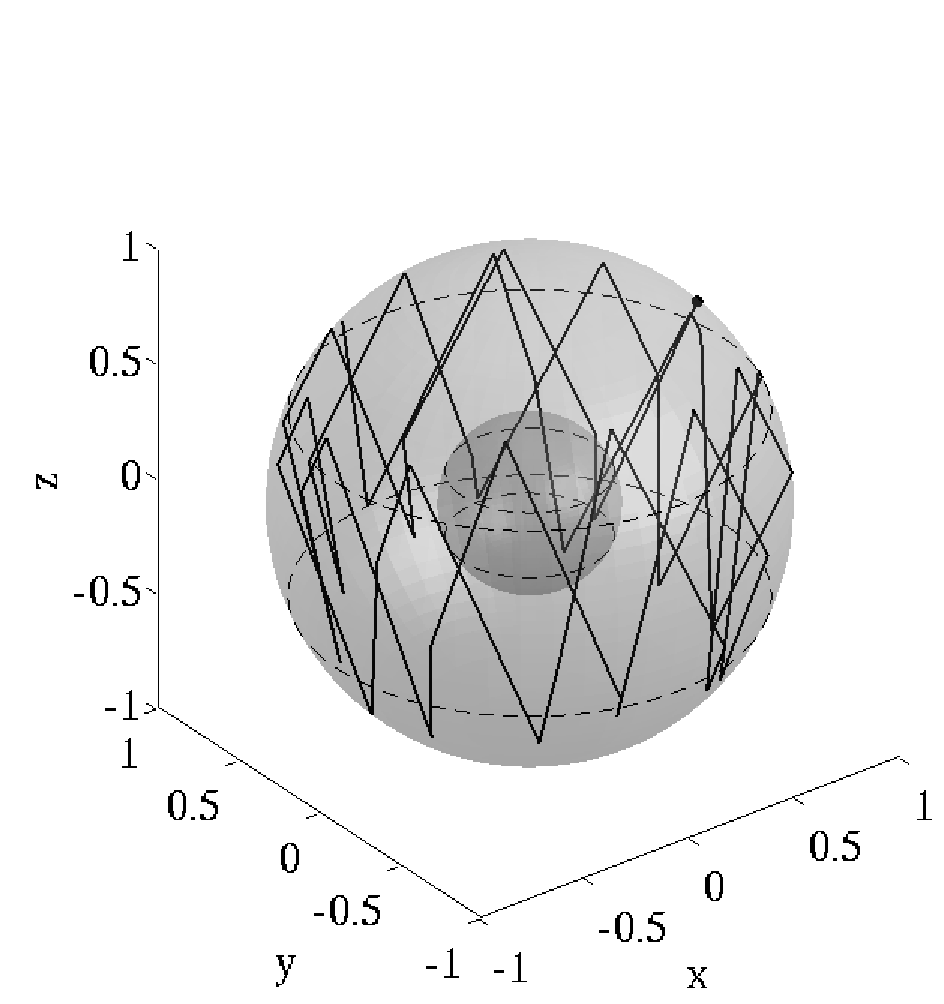}}
\caption{As figure \ref{fig:3d_erg}, example of a non converging trajectory, corresponding to star in figure \ref{fig:scan_xphi}a. Here $\eta = 0.35$, $\omega = 0.4051$, $x_0 = 0.8$, $\phi_0 = 1.8$ and $50$ reflections are depicted. \label{fig:eta35white}}
\end{figure}
\begin{figure}
\centerline{
a)
 \includegraphics[width=.3\linewidth]{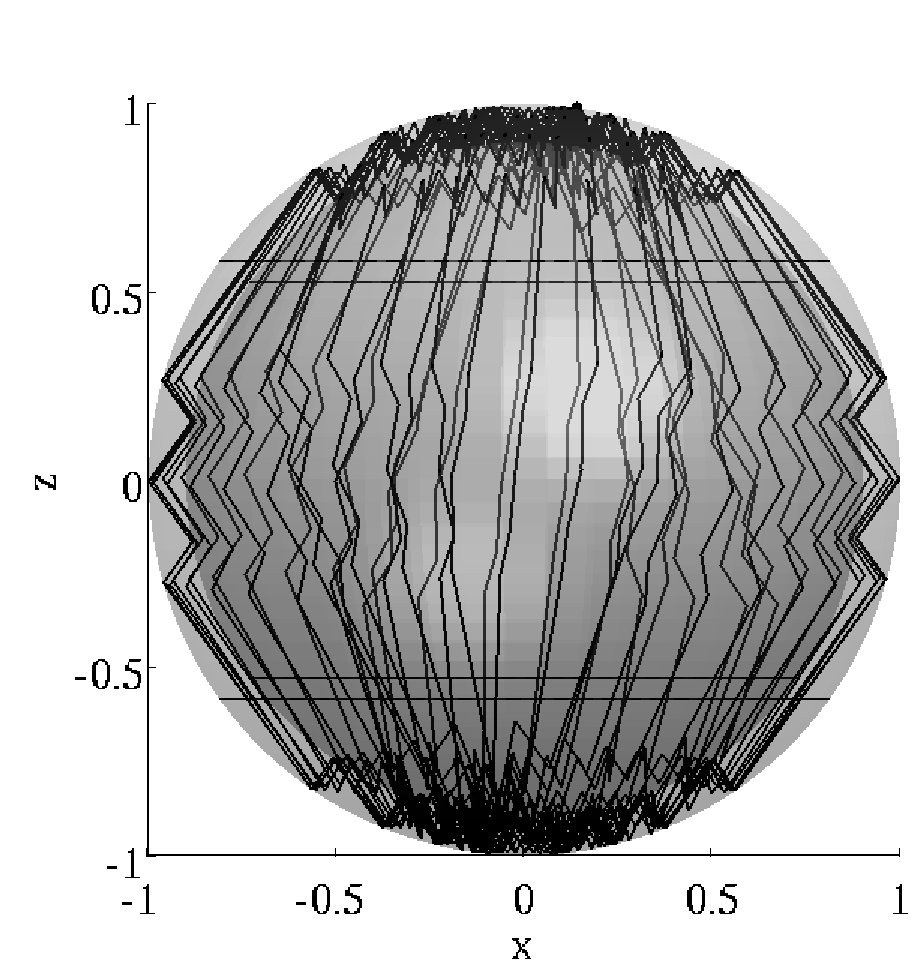}
b)
 \includegraphics[width=.3\linewidth]{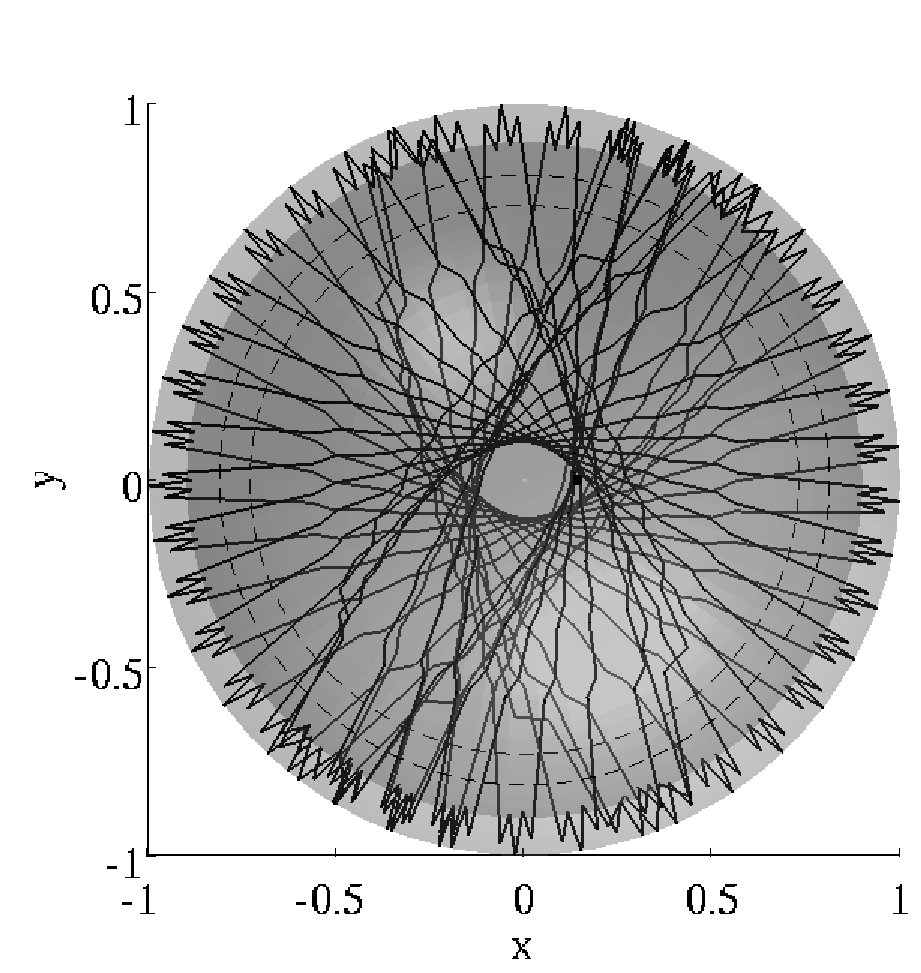}
c)
 \includegraphics[width=.3\linewidth]{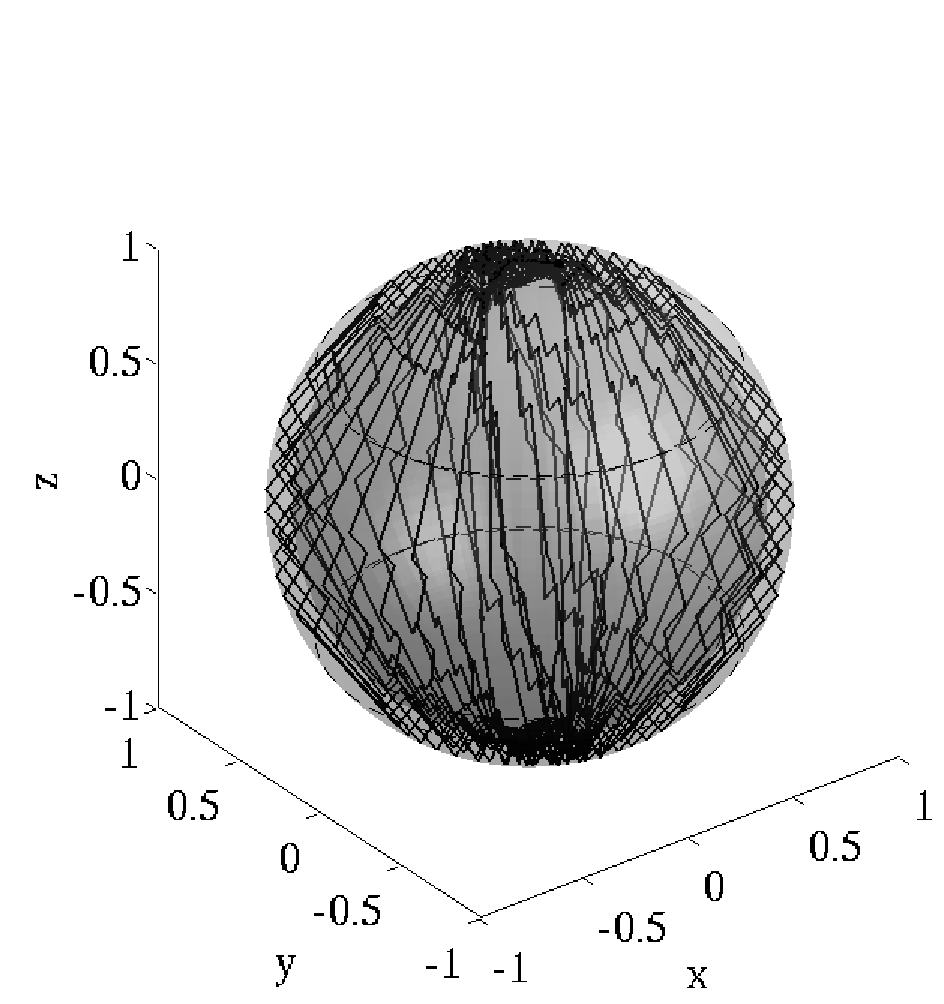}} 
\caption{As figure \ref{fig:3d_erg}, example of a non converging trajectory, corresponding to star in figure \ref{fig:scan_xphi2}b. Here $\eta = 0.9$, $\omega = 0.5842$, $x_0 = 0.139$ , $\phi_0 = -2.042$ and $1000$ reflections are depicted. \label{fig:eta9white}}
\end{figure}
As can be observed, for example, in figure \ref{fig:scan_xphi}a, even for a frequency value that gives rise to a meridional attractor, regions of the characteristic cone exist that are not subject to meridional trapping (white areas).
In figure \ref{fig:eta35white} the trajectory corresponding to the black star in figure \ref{fig:scan_xphi}a is displayed. This is representative of all the orbits in the white region on the right hand side of figure \ref{fig:scan_xphi}a, and it shows that these orbits do not interact with the inner sphere, that is, they behave as living in a full rotating sphere filled with fluid. As we expect from the known analytical solutions for this case, the full sphere does not support any singular orbits (attractors), and therefore the whole area of ``spherical'' orbits living in the shell coherently shows no trapping. 
We refer the reader to \S\ref{sec:sphere} for comments on three dimensional ray tracing in the limit case of the full sphere.
In this non-trapping area, the ray we are following within the three dimensional cone keeps on bouncing around the domain, experiencing subsequently focussing and defocussing reflections. 
It resembles the ``edge waves'' observed by \citet{Drijfhout2007} in the paraboloidal channel, with the only difference that here the trajectory, instead of being trapped around a single critical line, is now trapped in the equatorial belt, hugging both critical circles of the outer sphere.
As we can expect, the width of these non trapping areas reduces with increasing values of $\eta$ (see analogous white lobes on the right hand side of both figures \ref{fig:scan_xphi2}a and b): the bigger is the inner sphere, the more likely it is for the ray to touch it.

Interestingly, as we decrease the thickness of the shell, other white areas appear, according to a regular and fascinating pattern.
In these white regions we can now distinguish two different kinds of behaviour.
An example of non-converging orbits of the first type, corresponding to the black star in figure \ref{fig:scan_xphi2}a, is shown in figure \ref{fig:eta8white}. 
We can describe this kind of trajectories as ``polar'' edge waves, in analogy with the previous ``equatorial'' ones, and, consistently, ``polar'' edge waves are not periodic nor present any attracting power. 
However, surprisingly, ``polar'' edge waves do not seem to interact with the outer nor with the inner critical latitudes. 
It can be at first hypothesized that the ray would sense not only the gradient of the topography, but also the gradient of the fluid depth, and it could behave like an edge wave around this secondary ``critical'' latitude. This does not seem to be the case for ``polar'' edge waves, whose nature remains unexplained.
The second type of behaviour, displayed in figure \ref{fig:eta9white} (black star in figure \ref{fig:scan_xphi2}b), contrasting with the ``equatorial'' and ``polar'' edge waves, involves the whole fluid domain. The ray is never trapped onto a unique meridional plane, but it continuously and smoothly drifts in the zonal direction with no preferred direction of propagation.

Remarkably, all types of edge waves are not subject to meridional trapping, and therefore are invisible in a purely meridional (two-dimensional) ray tracing in spherical shell domain. It may be argued that these trajectories do not represent a physical solution in the domain, being the three-dimensional counterpart of the two-dimensional, annihilating, domain filling trajectories.
On the other hand, numerical results from \citet{Drijfhout2007} show the same structures, and it has been speculated they could provide an explanation to areas near the bottom of the ocean where enhanced internal wave activity is detected \citep{Horn1976}.
Moreover, the existence of zonally propagating modes could provide a rationale for the otherwise unexplained experimental results by \citet{Koch2012}, where only 16\% of the total energy measured in the wave field can be explained by purely meridional motion in a homogeneous, rotating spherical shell.
In analogy with electromagnetic phenomena as electron orbits, we could expect to observe in a real fluid only those trajectories that interfere constructively and show a periodic character, and close onto themselves after one azimuthal revolution around the domain. These kind of periodic trajectories would act as three-dimensional traps for rays, forced to travel endlessly along the respective periodic orbit (as in a three-dimensional attractor).
Promisingly, all kinds of edge waves show a regular pattern to a top view observer, but unfortunately no periodic edge wave has been observed so far, suggesting that this class of solution, if existing, does not posses any attracting properties.

It is fascinating, any way, how edge waves in the spherical shell reveal their strong dependency on the thickness of the domain.
As can be observed, comparing figures \ref{fig:scan_xphi2}a and \ref{fig:scan_xphi2}b, the two structures are basically the same. Increasing the value of $\eta$, in figure \ref{fig:scan_xphi2}b, we just witness an unfolding of the lobes already present in figure \ref{fig:scan_xphi2}a, and a flourishing of a crown of smaller white areas around the central one, resembling a fractal type of behaviour.

\subsection{Limit cases: the full sphere and the infinitely thin shell}
\label{sec:sphere}  

Analytical solutions to equation (\ref{eq:PE}) in a spherical geometry, completely filled with homogeneous or non homogeneous fluid,  are well known since the work by \citet{Bryan1889}, \citet{Friedlander1982a}, \citet{Friedlander1982b}, and \citet{Greenspan1968}, and, more recently, explicit solutions for the corresponding velocity field have also been derived by \citet{Zhang2001}.
Solutions are regular throughout the whole domain: three dimensional ray tracing analysis applied to the full sphere does not show indeed any  longitudinal inhomogeneity, and consequently neither two- nor three-dimensional attractors occur.
As anticipated in the previous section, because of its regularity, the full sphere could be a domain where (possibly) three-dimensional periodic trajectories appear, at frequencies corresponding to the known eigenvalues of the system. These periodic orbits would constitute the exact three-dimensional counterpart of the two-dimensional periodic trajectories, being representative of the regular modes existing in the sphere.

Orbits like the one depicted in figure \ref{fig:eta35white} seem appealing, but no three-dimensional periodic orbits have been found in the sphere, so far. They could either not exist, or they could be unstable, acting as repellors for the trajectories and in such a way prevent their observation. 
\citet{Dintrans1999}, using a combination of two-dimensional ray tracing and a three dimensional numerical model, have already suggested that pure inertial modes in a sphere could constitute an example of the association of regular modes to quasi-periodic (ergodic) orbits, breaking the usual association valid in two-dimensional domains, between closed (periodic) orbits and regular solutions.
However, we can not, in principle, exclude the existence of three-dimensional periodic orbits in the sphere, and consequently in the spherical shell as well, in which case the ray, because of its orientation, does not interact with the inner boundary (see, again, figure \ref{fig:eta35white}).
These orbits, if existent, would represent a set of exceptional non-annihilating edge waves in the fluid, and constitute an entire new class of waves, neglected so far in the purely meridional studies, because of their intrinsic three-dimensional nature.

The opposite scenario, the case of an infinitely thin shell ($\eta \rightarrow 1$) also deserves some special attention.
This case has been studied analytically in \cite{Stewartson1969, Stewartson1971, Stewartson1972}, where for the first time a ``pathological'' (singular) behaviour has been described as characterizing solutions in a rotating fluid shell.
Interestingly, frequency windows given by \cite{Stewartson1972} are easily retrieved as attractor frequency windows in the limit of a numerically infinitely thin shell, as already verified in the two dimensional meridional plane by \cite{Maas2007my}.
The singular nature of inertial wave solutions in shell geometries appears thus to bear its evidences at all values of $\eta > 0$.

\section{Discussion and conclusions}
\label{sec:discussion}
Three-dimensional internal wave ray tracing has here been systematically applied to rotating spherical shell geometries of arbitrary ratio between inner and outer radii.
In this work, the analysis is restricted to inertial waves only. Results for a homogeneous rotating shell are likely to be locally distorted in the presence of radial density stratification, but is assumed that the pure inertial problem summarizes already many difficulties and features of the gravito-inertial problem (with the exception of the turning surfaces, occurring only in the density stratified case), and findings in this partial case are supposed to have general applications.
We are conscious of the fact that a ray tracing study allows us to explore the behaviour of the inviscid linear solution only. 
However, even this partial perspective can be of help in improving our poor understanding of internal wave field dynamics in arbitrarily shaped enclosed fluid domains where, generally, no analytical (inviscid) solution is known, the spherical shell constituting a paradigmatic example of this category.
In spite of the azimuthal symmetry of the domain, in this work and differently from literature, ray motion is not restricted to the meridional plane only, but it is followed as it develops in a fully three-dimensional environment, allowing, in the first place, for a better representation of a local point source in the domain, and, secondly, for subsequent development of zonal inhomogeneities.

It is found that some frequency bands in the inertial wave range support meridional trapping of the rays. In these bands, internal wave ray trajectories, whose motion is initiated outside a meridional plane, are eventually trapped onto a meridional plane, from which they cannot escape any longer.
We call this meridional plane a ``meridional attractor'', in analogy with the two-dimensional internal wave attractors described in \citet{Maas1995}.  
 Once on a meridional plane, rays are subject to the occurrence of two-dimensional attractors, as it has been shown already  in \citet{Dintrans1999}, \citet{Rieutord2001}, and lately in \citet{Maas2007my}.
The fact that attractors in the shell act in a three-dimensional fashion comes with no surprise. 
In fact they are representative of the singular nature of internal wave field solutions in domains such as the spherical shell \citep{Bretherton1964, Stewartson1971, Stewartson1972}, and they comfortably emerge not only in the purely meridional, two dimensional representation of those solutions, but also when full, three-dimensional effects are taken into account.

If, on the one hand, the presence of meridional attractors justifies the use of ray tracing on meridional planes only, the present work on the other hand also points at the existence of zonally propagating waves in the inertial frequency range, whose trajectories could be traced here for the first time thanks to the adoption of a three-dimensional scheme.
These ray trajectories constitute a new and interesting class of possible solutions, so far neglected in purely meridional studies.
Even if the physical relevance of these type of orbits is not completely clear yet, they could help in the interpretation of some laboratory \citep{Hazewinkel2010a, Koch2012} and numerical \citep{Drijfhout2007} three-dimensional results.
It is worth noticing that trajectories that do not interact with the inner boundary constitute a subset within the zonally propagating edge waves. They can be thought of as trajectories belonging to the limit case of the full sphere, as the inner sphere is not sensed by the rays. 
One of the main questions remaining about these edge waves, both living in the spherical shell and in the sphere, concern three-dimensional orbit periodicity.
It is appealing to retrieve the usual association, valid in two-dimensional frameworks, that periodic orbits would represent regular solutions (the \textit{modes}) of the studied system, especially regarding the full sphere case, where eigenvalues are known.
It is, however, beyond the scope of this paper to address such a question.

In the present work, conditions for the occurrence of meridional attractors have been explored, motivated by possible astrophysical and geophysical applications of the results.
In fact, singular (attracting) type of solutions are supposed to play a role in diapycnal and angular momentum mixing, in regions where focussing reflections take place, and energy is localized to confined areas in the domain.

\subsection{Oceanographic implications}

Results presented in this study, obtained by means of ray tracing, are restricted to linear, inviscid fluid dynamics in three spatial dimensions and to a homogeneous rotating fluid, which clearly restricts their direct application to a real geophysical fluid. 
However, when it comes to geophysical applications, in the work of \citet{Broutman2004a} the ray approach has been analysed in a selection of case studies, where it has been generally recognized to provide a unique contribution to the understanding of spatial structures and spectra of atmospheric and oceanic internal waves (in the usual WKB approximation).
In the ocean, internal inertio-gravity waves are generated near ocean boundaries, at the surface by atmospheric perturbations, or over deep topographic features, by tidal forcing. These waves are observed to travel as confined energetic beams that can propagate through the ocean for thousands of kilometres \citep{Zhao2010a} and they constitute, by means of breaking and other small scale processes, one of the contributors to the deep ocean vertical turbulent diffusivity, necessary for maintaining the stratification and, over all, the global overturning circulation \citep{Wunsch2004}.
In case features as internal wave attractors occur in Nature, they could supposedly be responsible for strong energy focussing in specific locations in the interior of the fluid domain, possibly far from boundaries, with consequent local enhancement of wave breaking, mixing, and small scale processes, because of the regularization of the associated solution by means of enhanced viscous effects \citep{Bretherton1964, Maas2001, Harlander2007bb, Maas2007my}.
It is clear that various physical circumstances in planets, stars, oceans and atmospheres, may not always permit the numerous reflections that are needed by an attractor to develop. 
However, as already stated at the end of \S\ref{sec:3d}, it is not necessary to have a fully developed attractor to observe an increase of energy in a limited portion of the domain. 
This hypothesis has been confirmed in the three-dimensional laboratory experiments performed by \citet{Hazewinkel2010a}, and is at the base of any further speculations about the role of inertial waves (or, more generally, gravito-inertial waves) in the ocean, as well as in other media, and their possible interaction with the mean flow. 

In fact, density and angular momentum mixing due to internal wave breaking has  already been observed to generate a mean flow
 \citep{Maas2001, Tilgner2007, Swart2010, Morize2010, Sauret2010, Grisouard2012}.
This kind of mean flow generation process has been proposed as feeding mechanism for highly coherent (prograde) zonal currents (jets) in media, \citep{Maas2001, Maas2007my}, especially in the low latitude regions, where trapped wave solutions \citep{Stern1963} are generally focused onto periodic paths, leading to unstable regimes.
Local effects such as internal wave attractors and consequent angular momentum mixing could thus be at the basis of general phenomena, as the Equatorial Deep Jets, observed in all equatorial oceans (\citet{Firing1987} in the Pacific, \citet{Send2002} and \citet{Brandt2011a} in the Atlantic, \citet{Dengler2002} in the Indian Ocean) as well as in the atmosphere \citep{Galperin2004, Ogilvie2004}, or the variability in the rotation rate of rapid stars \citep{Balona1996}, phenomena whose forcing and maintenance have not yet been understood completely.

With the results presented in this work, attractor occurrence and related processes can now be understood afresh.
Not only are they valid when a perfectly axisymmetric forcing (and, consequently, wave motion) is present, but, due to the meridional focussing power of geophysical domains as a shell-like ocean or atmosphere, they are likely to take place even when point source and meridional inhomogeneities come into play, certainly a more realistic condition for a medium as our ocean.

 \vspace{0.5cm}
A. R. is supported by a grant from the Dutch National Science Foundation NWO.
 The authors gratefully acknowledge U. Harlander for the numerous and helpful comments on this work. 
 Thanks are also due to the anonymous referees for constructive suggestions and for improving lucidity of the manuscript.
%%%%%%%%%%%%%%%%%%%%%%%%%%%%%%%%%%%%%%%%%%%%%%%%%%%%%%%%%%%%%%%%%%%%%%%%%%%%%%%%%%
\appendix
%\section{}%[Appendix]{Three-dimensional algorithmic reconstruction of ray path in a spherical shell}
\label{app:shell_algo}
We discuss here the algorithmic reconstruction of the three dimensional path of an inertial wave beam in a spherical gap, subsequently reflecting on the curved boundaries of the convex (outer) sphere and of the concave (inner) sphere. 
The reflection of internal/inertial waves from a linearly sloping bottom was considered previously for plane waves in \citet{Phillips1963}, \citet{Phillips1966}, \citet{Greenspan1968}, \citet{Wunsch1968}, \citet{Wunsch1969}, \citet{Eriksen1982b}, \citet{Thorpe1997}, \citet{Thorpe2001} and, from a curved bottom, by \citet{Gilbert1989}. 
Here we follow the derivation in \citet{Phillips1963}, as proposed in \citet{Manders2004}. 
However, we will not use as framework of reference the plane defined by the incident and reflected rays, useful to study a single reflection alone \citep{Phillips1963}. 
When considering multiple, subsequent reflections, it is much more convenient to choose a reference framework having a fixed direction with respect to the restoring force acting in the system. Here we thus align the vertical with the rotation axis, and the horizontal frame will be reoriented at each reflection, the $x$-axis pointing in the direction of decreasing depth (outward).

Now consider we want to trace the behaviour of a perturbation at definite frequency $\omega$. This will travel along a double cone (one propagating upward and one propagating downward) of internal wave rays.
We will trace one single ray at a time, and it will be uniquely defined by the sign of its initial vertical velocity and three parameters: $\omega$, $\vec{x}_0$ and its initial horizontal direction $\phi_{0}$, measured anticlockwise with respect to the $x$-axis, which distinguishes it from other rays belonging to the same excited internal wave cone. 
Let the $n^{th}$ segment of the ray we are following be defined by a line $l_n$, and denote the time-derivative of the position along this ray, $\frac{d \vec{x}}{d t} = (\mathtt{u}, \mathtt{v}, \mathtt{w})$, which are the three components of the group-velocity vector. The group velocity will by definition be parallel to $l_n$.
It is assumed further that the vertical has been stretched according to the perturbation frequency so as to maintain the angle of the ray with the vertical fixed at 45\textdegree : 
\begin{equation}
 z' = z\frac{\omega}{\sqrt{1-\omega^2}}.
\end{equation}
This approach has general validity, and bears a strong simplification in the calculations, the only price being the compression or elongation of the spherical shell domain into an oblate ($\omega < 1/\sqrt{2}$) or a prolate spheroidal ($\omega > 1/\sqrt{2}$) shell. 
To facilitate the reader, all figures in this paper have been subsequently re-stretched back to the original geometry.
Moreover, in this study we are not interested in the group velocity magnitude \citep{Shen1975}, but in its direction only, therefore we can write without loss of generality
\begin{equation}
 \vec{c}_g = (\mathtt{w}\cos\phi_i,\mathtt{w}\sin\phi_i,\mathtt{w})
\end{equation}
 where 
\begin{equation}
\mathtt{w}=\pm 1
\end{equation}
according to the initial launching direction. 
Here $\phi_i$ represents the ``incoming'' horizontal direction of the ray approaching a boundary, after the $x$-axis has been reoriented, pointing outward at the location where $l_i$ intersects the boundary figure \ref{fig:reflection}.
As shown in \S\ref{sec:geom}, since the wave frequency does not change under reflection, the internal wave beam's angle with respect to the rotation axis (vertical) will also not change under reflection with the boundary: this is equivalent to requiring that the incident and reflected waves obey the same dispersion relation, equation (\ref{eq:dispersion}). 
However, the group velocity vector changes magnitude upon reflection from the bottom, because, in the cross-slope direction, rays change distance, as is clear from figure \ref{fig:reflection}a. This unusual property, typical of internal waves, and due to their anisotropic character, causes the wave number to change under reflection and energy to be transferred from one wave number to another.
In an inviscid fluid, at reflection, the boundary condition of vanishing normal flow of the group velocity (energy flow) has to be also satisfied. 
This requires that in the incident and in the reflected ray the group velocity component in the along-slope direction (along the $y$-axis in the rotated framework), $\mathtt{v}$, is unchanged, and is aligned with the isobaths.
No net group velocity at reflection means
\begin{equation}
 (\mathtt{w}_{r}\cos\alpha + \mathtt{u}_{r}\sin\alpha)=-(\mathtt{w}_{i}\cos\alpha + \mathtt{u}_{i}\sin\alpha)
\end{equation}
where we define $\alpha$ as the angle which the bottom makes with the up-slope directed x-axis, where subscriptions $r$ and $i$ denote the reflected and incident ray respectively.
With $s = |\nabla H| =\tan\alpha$, this can be simplified to
\begin{equation}
\mathtt{w}_{i}+\mathtt{w}_{r}=-s(\mathtt{u}_{i}+\mathtt{u}_{r}).
\label{eq:cond1}
\end{equation}
Since in this framework $\mathtt{v}$ is constant we can also write
\begin{equation}
\mathtt{v}_i^2 = \mathtt{w}_{i}^{2}-\mathtt{u}_{i}^{2}=\mathtt{w}_{r}^{2}-\mathtt{u}_{r}^{2} = \mathtt{v}_r^2
\end{equation}
that can be rewritten as
\begin{equation}
 (\mathtt{w}_{i}-\mathtt{w}_{r})(\mathtt{w}_{i}+\mathtt{w}_{r})=(\mathtt{u}_{i}-\mathtt{u}_{r})(\mathtt{u}_{i}+\mathtt{u}_{r})
\end{equation}
and therefore, with equation (\ref{eq:cond1}), $\mathtt{w}_{i}+\mathtt{w}_{r}$ can be eliminated:
\begin{equation}
 \mathtt{u}_{r}+s\mathtt{w}_{r}=\mathtt{u}_{i}+s\mathtt{w}_{i},
\label{eq:cond2}
\end{equation}
where we assume no vertical wall ($\mathtt{u}_{i}+\mathtt{u}_{r}\neq0$). The vertical wall case is treated later in this appendix.
From equations (\ref{eq:cond1}( and (\ref{eq:cond2}) $\mathtt{w}_r$ and $\mathtt{u}_r$ immediately follow
\begin{equation}
\left . \begin{array}{l}
 \mathtt{w}_{r}=\frac{-2s\mathtt{u}_{i}-(1+s^{2})\mathtt{w}_{i}}{1-s^{2}}\\
\mathtt{u}_{r}=\frac{(1+s^{2})\mathtt{u}_{i}+2s\mathtt{w}_{i}}{1-s^{2}}.
\end{array}
\right.
\label{eq:wrur}
\end{equation}
Together with the conditions $\mathtt{u}_{i,r}=\mathtt{w}_{i,r}\cos\phi_{i,r}$ and $\mathtt{v}_{i,r}=\mathtt{w}_{i,r}\sin\phi_{i,r}$, this implies an explicit relation between the incident and reflected angle:
\begin{equation}
 \sin \phi_r = \frac{(s^{2}-1)\sin\phi_i}{2s\cos\phi_{i}+s^{2}+1}
\end{equation}
(equation (\ref{eq:phir}) in \S\ref{sec:geom}) which conforms with the expression in \citet{Eriksen1982b}. 
The angle of reflection thus depends only on the angle of incidence of the ray, $\phi_{i}$, and on the bottom slope $s$. 
Regarding the sign of the vertical component of the group velocity $\mathtt{w}$, this is maintained when the bottom slope is locally supercritical (bottom slope larger than ray inclination), and reverses otherwise.
These transformations also apply at the flat surface ($s=0$), where they imply $(\mathtt{u}_r,\mathtt{v}_r,\mathtt{w}_r) = (\mathtt{u}_i,\mathtt{v}_i,-\mathtt{w}_i)$, meaning that the ray proceeds in the same horizontal direction while just reversing its vertical motion, and at a vertical wall ($s =\pm\infty$), where $(\mathtt{u}_r,\mathtt{v}_r,\mathtt{w}_r) =(-\mathtt{u}_i,\mathtt{v}_i,\mathtt{w}_i)$.
It is possible to investigate the focussing or defocussing nature of a single reflection looking at the changes in the group velocity magnitude upon reflection. 
Because of energy conservation, after a focussing reflection, and the consequent decrease in wave length (see figure \ref{fig:reflection}b for example: rays propagate according to the arrows), group velocity magnitude must decrease. Conversely, after a defocussing reflection (see again figure \ref{fig:reflection}b for example, but with rays propagating in a direction opposit to the arrows), group velocity magnitude must increase. 
In the first case (focussing reflection) initial parameters are $\mathtt{w}_i=-1$, $\mathtt{u}_i=1$  (case of normal incidence, $\mathtt{v}_i=0$) and $s<1$. Applying equation (\ref{eq:wrur}) we obtain $|\mathtt{w}_r|<|\mathtt{w}_i|$ and $|\mathtt{u}_r|<|\mathtt{u}_i|$.
Knowing $|\mathtt{v}_r|=|\mathtt{v}_i|$, $|\mathbf{c}_{g,r}|<|\mathbf{c}_{g,i}|$ follows, as expected. 
In the second case (defocussing reflection) initial parameters are $\mathtt{w}_i=-1$, $\mathtt{u}_i=-1$ and $s<1$. Applying equation (\ref{eq:wrur}) we obtain the reverse result, with $|\mathbf{c}_{g,r}|>|\mathbf{c}_{g,i}|$ as expected. 
Interestingly, if we approach a critical reflection ($s=1$) in the first case, $|\mathbf{c}_{g,r}| \rightarrow 0$, this means that all rays are reflected along one single line approximately coinciding with the sloping wall itself.
But if we approach a critical reflection in the second case, $|\mathbf{c}_{g,r}| \rightarrow \infty$, this means the incident ray is already travelling approximately along the slope, and two neighbouring rays could, in this limit, reflect infinitely far from each other.
The geometrical mechanism of repeated reflections of internal waves can thus be studied provided at each reflection point we realign the $x$-axis with decreasing depth prior to application of the reflection laws. 

Given the dimensionless topography, which defines the domain, determined by the outer sphere, $z = h_{out}(x,y)=\pm\sqrt{1-x^{2}-y^{2}}$, and the inner sphere,  $z = h_{in}(x,y)=\pm\sqrt{\eta^{2}-x^{2}-y^{2}}$, we choose an initial location $(x_{0},y_{0},z_{0})$, on the outer or on the inner sphere, an initial horizontal direction, $\phi_{0}$ (one ray on cone), and the initial sign of the vertical group velocity component $\mathtt{w}_{0}= +1$ or $-1$, in accordance with the location of the initial point. The magnitude  $|\mathtt{w}_0|$, not relevant to our problem, has arbitrarily been set equal to $1$ here, and $|\mathtt{w}_i|$ is reset to $1$ prior to each reflection.
Consistently, we thus do not reconstruct algorithmically the changes in wave amplitude along the ray path. This problem has been investigated in detail in the work by \citet{Shen1975}, in the specific case of oceanic or atmospheric propagating waves.
We remark that the topography is described in stretched coordinates, so that, as already mentioned, the vertical inclination of the ray is always $\theta = 45$\textdegree.
It is then possible to determine iteratively the subsequent intersections with bottom and surface, $(x_{n},y_{n},z_{n})$, as well as the horizontal angle $\phi_{n}$ and group velocity $\mathtt{u}_n,\mathtt{v}_n,\mathtt{w}_n$ (of which only the direction is relevant) applying the following algorithm.
\begin{enumerate}

 \item According to the launching point and the direction of propagation, determine the proper time $t$ of intersection between the ray and the inner or the outer sphere:
\begin{equation}
 (x,y,z)=(x_{n},y_{n},z_{n})+\mathtt{w}_{n}t(\cos\phi_{n}, \sin\phi_{n},1).
\end{equation}
Care is needed in order to reject all trajectories passing through the core of the shell.
Denote that point by $(x_{n+1},y_{n+1},z_{n+1})$. At the time of intersection $t_{n+1}$, $z_{n+1}= h_{in,out}(x_{n+1},y_{n+1})$, while $\mathtt{w}_{n}t_{n+1}=h(x_{n+1},y_{n+1})-z_{n}$, where $h$ is intended from now on to be the intersection with the appropriate boundary (inner or outer shell) and the appropriate sign (northern or souther hemisphere). 
Then the horizontal coordinates $(x_{n+1},y_{n+1})$ follow from simultaneous solution of the equations
\begin{equation}
 x_{n+1}=x_{n}-(h(x_{n+1},y_{n+1})+z_{n})\cos\phi_{n}
\end{equation}
\begin{equation}
 y_{n+1}=y_{n}-(h(x_{n+1},y_{n+1})+z_{n})\sin\phi_{n}
\end{equation}

\item Determine the local gradient of the bottom: 
\begin{equation}
 \nabla h = (h_{x}(x_{n+1},y_{n+1}),h_{y}(x_{n+1},y_{n+1})),
\end{equation}
which leads to slope $s=|\nabla h|$ and direction $\sigma=\tan^{-1}(h_{y}/h_{x})$.

\item Determine the along-slope velocity component 
\begin{equation}
 \mathtt{v}_{n+1}=\mathtt{v}_{n}=\mathtt{w}_{n}\sin(\phi_{n}-\sigma)
\end{equation}
and the cross-slope velocity component 
\begin{equation}
 \mathtt{u}_{n}=\mathtt{w}_{n}\cos(\phi_{n}-\sigma),
\end{equation}
from which follows
\begin{equation}
 \mathtt{u}_{n+1}=\frac{(1+s^{2})\mathtt{u}_{n}+2s\mathtt{w}_{n}}{1-s^{2}}
\end{equation}
and the vertical velocity component
\begin{equation}
\mathtt{w}_{n+1}=\frac{-2s\mathtt{u}_{n}-(1+s^{2})\mathtt{w}_{n}}{1-s^{2}}.
\end{equation}
\item The angle $\phi_{n}-\sigma$ is the angle of incidence $\phi_{i}$ with respect to the local up-slope direction.
From the horizontal velocity components in the cross- and along-slope direction follows the new direction with respect to the original 
frame of reference
\begin{equation}
 \phi_{n+1}=\tan^{-1}(\mathtt{v}_{n+1}/\mathtt{u}_{n+1}) + \sigma
\end{equation}
and the algorithm can be iterated.
\end{enumerate}
%%%%%%%%%%%%%%%%%%%%%%%%%%%%%%%%%%%%%%%%%%%%%%%%%%%%%%%%%%%%%%%%%%%%%%%%%%%%%%%%%%
\bibliographystyle{jfm}
\bibliography{/user/rabitti/Desktop/scritti_vari/references/library,/user/rabitti/Desktop/scritti_vari/references/library_MY}

\end{document}